\documentclass[manuscript]{acmart}

\usepackage{geometry} 
\usepackage{multirow}
\usepackage{longtable}
\usepackage{booktabs}
\usepackage{array}
\usepackage{xcolor}
\usepackage{colortbl}

\definecolor{rowgray}{gray}{0.95}

\AtBeginDocument{%
  }

\setcopyright{none}
\copyrightyear{}
\acmYear{2026}
\acmDOI{XXXXXXX.XXXXXXX}
\acmConference[CSCW '26]{PACM on Human-Computer
Interaction}{2025}{TBD}
\acmISBN{}

\begin{document}

\title[LLMs in Wikipedia]{LLMs in Wikipedia: Investigating How LLMs Impact Participation in Knowledge Communities}


\author{Moyan Zhou}
\email{zhou0972@umn.edu}
\orcid{0000-0002-5420-6587}
\affiliation{
    \institution{University of Minnesota}
    \city{Minneapolis}
    \state{Minnesota}
    \country{USA}
}

\author{Soobin Cho}
\email{soobin30@uw.edu}
\orcid{0000-0002-4832-208X}
\affiliation{
    \institution{University of Washington}
    \city{Seattle}
    \state{Washington}
    \country{USA}
}

\author{Loren Terveen}
\email{terveen@umn.edu}
\orcid{0000-0002-8843-4035}
\affiliation{
    \institution{University of Minnesota}
    \city{Minneapolis}
    \state{Minnesota}
    \country{USA}
}


\renewcommand{\shortauthors}{}

\begin{abstract}
Large language models (LLMs) are reshaping knowledge production as community members increasingly incorporate them into their contribution workflows. However, participating in knowledge communities involves more than just contributing content - it is also a deeply social process. While communities must carefully consider appropriate and responsible LLM integration, the absence of concrete norms has left individual editors to experiment and navigate LLM use on their own. Understanding how LLMs influence community participation is therefore critical in shaping future norms and supporting effective adoption. To address this gap, we investigated Wikipedia, one of the largest knowledge production communities, to understand 1) how LLMs influence the ways editors contribute content, 2) what strategies editors leverage to align LLM outputs with community norms, and 3) how other editors in the community respond to LLM-assisted contributions. Through interviews with 16 Wikipedia editors who had used LLMs for their edits, we found that 1) LLMs affected the content contributions for experienced and new editors differently; 2) aligning LLM outputs with community norms required tacit knowledge that often challenged newcomers; and 3) as a result, other editors responded to LLM-assisted edits differently depending on the editors' expertise level. Based on these findings, we challenge existing models of newcomer involvement and propose design implications for LLMs that support community engagement through scaffolding, teaching, and context awareness. 

\end{abstract}




\maketitle

\section{Introduction}
Knowledge production refers to the process of creating, maintaining, and propagating knowledge \citep{alavi2001knowledge}. The laborious process has traditionally required contributors to carefully gather, verify, and synthesize information. But the advent of large language models (LLMs) has transformed this reality, as LLMs enable users to generate fluent and well-structured content through natural language prompts \citep{brachman2024knowledge, shao2024assisting}. Prior work has indicated the usefulness of LLMs as writing assistants and imagined the potential of LLMs in knowledge production \citep{wasi2024llms, lee2024design, aubin2024llms, shao2024assisting}. 

However, participation in knowledge communities involves more than writing fluent or well-structured paragraphs. It is also highly epistemic (e.g., factuality, balance) and social \citep{bryant2005becoming, ren2023did} (e.g., aligning with norms \citep{butler2008don, beschastnikh2008wikipedian}, engaging with other community members \citep{maddock2017talking, ren2023did}). There is a mismatch between surface-level fluent content produced by LLMs and the unique epistemic and social needs of knowledge communities, such as knowledge integrity \citep{ford2023chatgpt}. Furthermore, Wikipedia, one of the largest knowledge production platforms \citep{ayers2008wikipedia, ren2023did} has not yet reached a consensus or developed norms around LLMs \citep{ayers2008wikipedia, ren2023did}, which could exacerbate these challenges. As a result, individual editors must interpret, navigate, and experiment with appropriate and responsible usage of LLMs within the community. 

The tension between surface-level content generation and unique epistemic and social needs raises important questions about how LLMs are shaping community participation in knowledge communities, critical to support the credibility and sustainability of these knowledge communities. This is especially true given the rise of LLM-generated content in new articles observed in Wikipedia \citep{brooks2024rise}, indicating that editors have already incorporated LLMs during their contributions. Our study aims to understand the use of LLMs by knowledge community members. Specifically we ask: 

\begin{itemize}
\item {\textbf{RQ1}: How does using LLMs influence the ways editors contribute content? }
\item {\textbf{RQ2}: What strategies do editors leverage to conform to community norms for LLM-assisted contributions? }
\item {\textbf{RQ3}: How do other editors in the community respond and engage with LLM-assisted contributions? }
\end{itemize}

Through semi-structured interviews with 16 participants who have used LLMs to edit Wikipedia, we unveil an expertise-based participation divide between editors. Experienced editors enhance their participation through LLMs, as they expand the range of their contributions in Wikipedia, leverage strategies to align LLM-generated content with community norms, and thus receive positive responses from other members. In contrast, LLMs raise the demands for new editors to participate in the community. LLMs lower the barriers for entry, and new editors tend to rely on LLMs to fill in their knowledge gaps. However, LLMs compel them to make judgments about AI-generated content, which requires skills they have not yet developed. This challenges them to produce high quality content, leading to rejections from others. 

Based on our findings, we discuss how the participation paradox for newcomers, namely lower barriers to entry yet higher demands to participate, arises from LLM usage, breaking the trajectory of Legitimate Peripheral Participation (LPP). LLMs interrupt learning pathways for newcomers, thus new editors miss opportunities to gain wiki skills gradually. At the same time, LLMs blur the boundaries between peripheral and central tasks and push newcomers from peripheral participants to editorial judges, which requires more central wiki skills. Building on our findings, we then propose design implications to mitigate the participation gap LLMs can lead to by fostering newcomers’ learning process and supporting editors with varying expertise in knowledge production platforms. 

We situate our work within the broader empirical HCI/CSCW research on human AI collaboration and communities of practices \citep{mackenzie2024human, oulasvirta2016hci, erickson2000social}, particularly work around how tools reshape work \citep{norman1991cognitive, dillahunt2015promise}, how individuals align to norms and boundaries \citep{hsieh2023nip, halfaker2013rise, kiene2016surviving}, and how community respond and adapt to technology changes \citep{smith2020keeping, kuo2024wikibench}. By examining how editors integrate LLMs into their practices, we made the following contributions: 
\begin{itemize}
\item {We surfaced the invisible dynamics of how editors with different levels of experience adopt, adapt, and respond to LLMs in knowledge communities in practice; }
\item {We extend Legitimate Peripheral Participation by demonstrating that LLM usage interrupts newcomers' traditional learning pathway; }
\item {We informed the design of future LLMs for knowledge communities, highlighting the need for AI tools to scaffold, teach, and be context aware to support community participation. }
\end{itemize}

\section{Related Work}
\subsection{Human AI collaboration in writing}
Knowledge production involves writing, summarizing, and drafting information in textual format. To contextualize this process, we surveyed the literature in the domain of AI-assisted writing and identified patterns of human-AI collaboration.

Generative AI supports writing across a wide range of domains, from personal writing such as diary entries \citep{kim2024diarymate}, to creative writing including stories \citep{biermann2022tool, yuan2022wordcraft, lee2022coauthor, stefnisson2018mimisbrunnur, chung2022talebrush}, fictions \citep{yang2022ai}, metaphors \citep{gero2019metaphoria}, poetry \citep{chakrabarty2022help}, and screenplays \citep{tang2025understanding}, as well as academic writing \citep{lee2022coauthor, luther2024teaming, dhillon2024shaping}. As LLMs continue to advance, more domains will leverage their capabilities to enhance the writing process. 

In academic writing, which closely aligns with our focus on knowledge production, researchers have developed and evaluated tools to assist with various stages of the writing process. For example, Sparks \citep{gero2022sparks} was implemented to inspire connections between scientific concepts. A quantitative analysis of 14 million abstracts of published papers over 14 years revealed that at least 10\% of 2024 abstracts were likely LLM-assisted, indicated by the suspicious increase in the usage of certain words \citep{kobak2024delving}. Several studies \citep{agarwal2024litllms, tang2024llms} evaluated the potential of LLMs to conduct literature reviews. Additionally, Radensky et al. \citep{radensky2024let} introduced a plan-draft-revise workflow and designed a writing assistant for scientists to draft blog posts for their research papers. 

Researchers have also investigated how users perceive and interact with these AI tools. Users often perceive generative AI not merely as a passive writing aid \citep{jin2025high}, but an idea generator and active collaborator \citep{yang2022ai, dhillon2024shaping} in content creation. Engagement with LLMs differs significantly depending on the use case. For example, users engage in back and forth conversations with Sparks for translating concepts, while they use it more independently in longer sessions for inspiration and perspective taking \citep{gero2022sparks}. Implicit users who were less specific about their goals tended to search for new ideas, while explicit writers sought precise content to incorporate in their writing \citep{yang2022ai}. Similarly, ChatGPT was used to support users’ information seeking behaviors, rather than completing tasks for them \citep{luther2024teaming}. Writers may modify or directly accept the generated text \citep{yang2025modifying}, and doctoral students iteratively interact with GAI assistants in reading, copying, pasting and shaping content in the writing process \citep{nguyen2024human}. 

Building on this body of prior work, we examined how knowledge contributors use and interact with LLMs. Our findings extend existing research by revealing use cases within the knowledge production process and identifying similar user behaviors. Most importantly, we found expertise influences their ability to incorporate LLM-generated content in their own writing. As a result, their participation in the knowledge community is affected. 

\subsection{Knowledge production in Wikipedia}
In Wikipedia, participating in the collaborative knowledge work entails three major aspects: 1) contributing content 2) engaging with other editors and 3) enforcing norms. 

\subsubsection{Content}
The central component of knowledge work on Wikipedia focuses on the content or knowledge itself. Rooted in a range of intents such as adding supporting evidence and removing existing information \citep{rajagopal2022one}, editors perform insertions, deletions, modifications, and relocations \citep{daxenberger2012corpus, daxenberger2013automatically}. Such adjustments to articles target different elements in Wikipedia articles, including texts, links, references, and templates \citep{daxenberger2013automatically, daxenberger2012corpus}. Yang et al. \citep{yang2016edit, yang2016did} categorized edit actions into two larger groups: meaning-preserving edits and meaning-changing edits. While meaning-preserving edits include modifications such as paraphrasing, spelling/grammar, and relocation, meaning-changing edits contain insertions, deletions and some level of modifications. Built on 14-label taxonomy from \citep{yang2016edit}, Ruprechter et al. \citep{ruprechter2020relation, ruprechter2020relating} derive 3 super-labels: content as modification to actual information on the article page, format without changing of meaning on text, and WikiCotent such as processing tags and vandalism fighting.

Multilingual edits across language editions \citep{hale2014multilinguals} are important for knowledge production, especially on Wikipedia, because of (1) the nature of more than 300 language editions in Wikipedia \citep{roy2022information}, and (2) content coverage and language gaps that contributors aim to bridge \citep{redi2020taxonomy}. As the quality \citep{roy2022information} and quantity of articles across Wikipedia language editions vary diversely \citep{laxstrom2015content, wulczyn2016growing}, multilingual editors play an essential role in bridging and filling the gaps from one language edition to another \citep{hale2014multilinguals, wulczyn2016growing, kim2016understanding}. 

\subsubsection{Community}
As editors gain experience on Wikipedia, their focus often shifts from individual content edits to active engagement with the editor community \citep{bryant2005becoming}. The need for coordination increases, as social interactions with other editors help seek internal consensus \citep{beschastnikh2008wikipedian} (e.g., structure of the page \citep{ren2023did}), resolve conflicts \citep{kittur2010beyond, maddock2017talking}, especially when they work on the same articles \citep{ren2023did}. Editors engage in social interactions \citep{liu2011does} in talk pages and Wikiproject groups \citep{ren2023did}. Though a small group of editors participate in article talk page discussions \citep{kittur2008harnessing}, editors use both article talk pages and user talk page \citep{kittur2010beyond} for diverse purposes, such as requests or suggestions for editing coordination, requests for information, and references to vandalism \citep{yasseri2012dynamics}. Wikiprojects establish a sense of belonging for editors as they socialize with other similar-minded editors \citep{forte2012coordination, ren2023did}.

\subsubsection{Policies and Guidelines}
On top of content and community, an editor is expected to follow community norms. Wikipedia policies and guidelines are community norms that support collaborative work \citep{beschastnikh2008wikipedian} and maintain high quality content and credibility of Wikipedia \citep{priedhorsky2007creating}. These policies define not only the standards for allowed content (e.g., Neutral point of view \citep{wikipedia_npoV}, Verifiability \citep{wikipedia_verifiability}, Notability \citep{wikipedia_notability}), but also prescribe expected behaviors among community members (e.g., civility, dispute resolution, and no personal attacks) \citep{suh2009singularity}. Butler et al. \citep{butler2008don} identified their roles such as signals to external organizations protecting Wikipedia’s reputation and outside attack. They also represent collective rational efforts to organize, coordinate for consistent and reliable contributions. By easing the pains of seeking consensus \citep{beschastnikh2008wikipedian}, these rules play an essential role as boundary objects \citep{velt2020translations} to the daily operation in Wikipedia. 

In summary, content, community, and norms inspire and guide our research questions, as we reflect that participation in Wikipedia consists of these aspects. They also serve as a basic framework to inform our interview questions, in correspondence to knowledge contributors’ edits and contribution workflows.

\subsection{Tool-assisted editing in Wikipedia}
Editors and researchers have developed tools to ease the complexity of contributing to Wikipedia. In this section, we mainly reviewed semi-automated bots, and noted more recent AI/ML tools. 

\subsubsection{Semi-automated bots}
Bots play a vital role in Wikipedia’s collaborative editing system \citep{geiger2009social} and have been extensively studied in the literature \citep{tsvetkova2017even, cosley2007suggestbot, geiger2013levee, geiger2010work}. Bots are designed to automate repetitive or large-scale tasks \citep{tsvetkova2017even}, and most of the early bots are implemented by scripts. For instance, HostBot invites new editors to socialize in Q\&A forums \citep{morgan2018evaluating, morgan2013tea}. Huggle and Twinkle fight vandals and unconstructive edits \citep{geiger2010work}. Rambot adds data into country and city articles \citep{halfaker2012bots}. Zheng et al. \citep{zheng2019roles} classified bots’ roles into 9 categories: generator, fixer, connector, tagger, clerk, archiver, protector, advisor, and notifier. 

While bots protect and patrol high quality knowledge in Wikipedia \citep{geiger2013levee}, they can also cause unintended consequences. In particular, vandal fighter bots negatively impact newcomers’ retention, as they revert contributions from good faith editors that may appear suspicious \citep{zheng2019roles, halfaker2013rise}. This reflects a broader dynamic in human bot collaboration. Editors not only develop, approve \citep{geiger2017operationalizing}, operate, and maintain bots \citep{zheng2019roles, ren2023did}, but are also influenced by bots' behaviors and work alongside them. For example, human editors check and correct the information by bots \citep{niederer2010wisdom}. In the context of vandal fighting, bots detect potential vandals and put them in queues. Editors can then decide and authorize bots to revert edits and leave warning messages. In more severe cases, bots can notify administrators when users accumulate multiple warnings \citep{geiger2010work}. 

Because bots closely support the editing process, they are generally well accepted \citep{clement2015interacting}. In fact, they are not viewed as simple tools, but social actors \citep{geiger2018lives} that shape collaborative outcomes. Through the evolving collaboration dynamics, humans and bots maintain the values and social structures of Wikipedia together \citep{geiger2010work}.

Notably, emerging AI/ML powered tools are developed not only to assist content editing, but also to support tasks related to editing. Some tools \citep{kumarana2011wikibhasha, bronner2012cosyne, laxstrom2015content} focus on multilingual contributions. Some tools are designed to support tasks related to editing, while others make suggestions or improve content quality. For example, SuggestBot \citep{cosley2007suggestbot} and other recommendation systems \citep{wulczyn2016growing} make edit recommendations to editors. ORES \citep{halfaker2020ores} evaluates the quality of articles. Edisum \citep{vsakota2024edisum} generates informative edit summaries. Wikimedia Foundation recognizes the potential of AI tools, and is developing strategies to continue invest in AI tools that support the editing process and improve editors' experience \citep{wikimedia}.

Building on prior work, we identify interaction patterns that differ from observed in traditional bot usage. Unlike traditional tools that conform to Wikipedia norms, LLMs bypass policies and guidelines and require editors to make normative decisions, which challenge new editors' participation.

\section{Methods}
The researchers collaboratively designed a semi-structured interview protocol, which included 3 sections: 1) general context about contributions, 2) experience of using LLMs in Wikipedia, and 3) vision for human AI collaboration in knowledge production. After receiving an exemption from the Institutional Review Board at the University, the first author conducted pilot studies with two graduate students and refined the interview questions. Data from pilot studies were excluded from data analysis. 

\begin{table}[h!]
\centering
\renewcommand{\arraystretch}{1.3} 
\begin{tabular}{@{}clp{1.5cm}p{1.5cm}p{2cm}p{2cm}p{1.5cm}p{2.5cm}@{}}
\toprule
\textbf{ID} & \textbf{Tenure} & \textbf{\# of Edits} & \textbf{Frequency} & \textbf{LLM models} & \textbf{Gender} & \textbf{Age} & \textbf{Race} \\ \midrule
P01 & 0-2 years	 & \textasciitilde 100 & monthly & ChatGPT, Gemini, You.com  & Male	& 35-44 & White/Caucasian              \\
P02 & 2-5 years	 & 10k+ & daily& ChatGPT & 	&  &    \\
P03 & 5+ years & 21k+  & daily & ChatGPT, Grok	& Male	& 55+ & White/Caucasian               \\
P04 & 5+ years & 2k+ & weekly & ChatGPT &  & 25-34 & White/Caucasian            \\
P05 & 5+ years & 50k+ & daily & ChatGPT & Male & 25-34 & White/Caucasian        \\
P06 & 5+ years & 1k+	& weekly & 	ChatGPT	& Male	& 35-44	& Avropoid-Caucasian/Azerbaijanian turkish        \\
P07 & 5+ years & 159K+	& daily	& ChatGPT	& Male & 	45-54 &	Middle Eastern/Arab          \\
P08 & 5+ years	& 12k+	& daily	& ChatGPT	& Non-binary	& 18-24	& Asian        \\
P09 & 0-2 years	& 5.3K+	& daily	& ChatGPT; Claude & 	Male & 25-34 & 	White/Caucasian        \\
P10 & 2-5 years	& &	monthly &	ChatGPT	& Male	& 18-24	& Black or African American        \\
P11 & 	5+ years	& \textasciitilde 10	& yearly	& ChatGPT	& & &      \\
P12 & 5+ years & 55k+ & weekly & ChatGPT; Claude; Llama & Prefer not to answer &35-44	& White/Caucasian     \\
P13 & 2-5 years	& 4k+	& daily	& ChatGPT; Grok &Male	& 18-24 & Bengali               \\
P14 & 0-2 years & 200+	& & & & & 			\\
P15 & 0-2 years & 200+	& & & & & 		\\			
P16 & 5+ years & 83k+ &monthly & ChatGPT & Male & 35-44 & White/Caucasian      \\ \bottomrule
\end{tabular}
\caption{Participant summary, adopted from \citep{smith2020keeping}. Some fields are left empty due to incomplete responses or participants’ preference.}
\label{tab:participants}
\end{table}

\subsection{Recruitment and Participants}
We recruited participants through multiple channels. We posted our interview invitation on the WikiMedia project page. At the same time, we searched for editors who potentially used LLMs and sent invitations to their talk pages or via email if they had enabled that feature. Lastly, we leveraged snowball sampling, asking referrals from our participants. Editors who were interested in the interview study filled out a survey about basic usage of LLMs, experience on Wikipedia, contact and demographic information. We then sent emails to schedule the interview on Zoom, along with a consent form. Several participants indicated discomfort with video/audio recording. Thus, we offered the option to participate in the interview through written text (email).

As a result, we conducted 16 interviews, of which 5 were conducted via email. We began to observe data saturation after finishing the 13th interview. Thus, we stopped recruiting more participants after completing 16 interviews, which was consistent with the average sample size for qualitative research at a top HCI conference \citep{caine2016local}. Participants did not receive monetary compensation, and their demographic information along with their expertise in Wikipedia is listed in Table \ref{tab:participants}.

\subsection{Interview Procedure}
Consent was acquired from each participant before the interview began. Before the interview, we thanked them for their participation and noted the privacy considerations - the interview would be recorded only for data analysis purposes; video is not required; participation would remain anonymous. As each participant fully understood and agreed, we began the interview and followed the interview protocol. 

First, we briefly introduced the goal of the study. We then asked participants about their contribution process such as typical tasks, general workflow, and common practices or guidelines to understand the context. Next, we asked participants to think of a specific use case of LLMs in their contribution, and asked relevant questions such as interactions with LLMs. Then, we asked them to envision the future for human AI collaboration in knowledge work and ideal interactions with LLMs. We concluded the interviews by inviting participants to share any additional information. In the semi-structured interview, we allowed participants to freely share their experiences and asked follow-up questions.

\subsection{Data Analysis}
Our 11 recorded interviews ranged from 42 minutes 22 seconds to 1 hour 17 minutes 29 seconds, with an average of 1 hour 3 minutes 18 seconds. After transcribing the interviews, we conducted an inductive thematic analysis approach \citep{clarke2017thematic}. We used ATLAS.ti, a popular software for qualitative data analysis. First, the authors open coded three interviews together to establish a shared standard for open coding. Then, we asynchronously open coded the remaining interviews. We ended up with 1524 codes, for example, “P05 - LLMs has no sources”. 

After that, the authors collaboratively grouped the codes based on their similarity and relevance in Miro. Themes and sub-categories emerged through this iterative process, and the authors discussed and reached agreement for the name of the themes and sub-categories. For example, the code mentioned above was put under the sub-category ``no sources," and the theme ``violates Wikipedia policies."

\section{Results}
Our results surfaced how knowledge contributors leverage and interact with generative AI (see Table \ref{tab:uses} for LLM use cases), showing that LLM usage influences participation in knowledge production communities in three ways: contributing content, enforcing norms, and other members' engagement. Across these dimensions, we identified a participation divide between editors, shaped by differences in expertise. 

\begin{table}[ht]
\centering
\begin{tabular}{>{\bfseries}l>{\raggedright\arraybackslash}p{0.25\textwidth}>{\raggedright\arraybackslash}p{0.6\textwidth}>{\raggedright\arraybackslash}p{0.0\textwidth}}
\toprule
\textbf{Category} & \textbf{Use cases} & \textbf{Highlights} \\
\midrule
 \multirow{4}{*}{\centering Generate} &  \cellcolor{gray!30} Generating examples & \multirow{4}{*}{\parbox{9cm}{ \textit{Generating practical examples for theoretical concerns.} (P14) \newline \textit{I’ve been learning. So, I need a guide... If there is anything that could guide me for better content.} (P15)}} \\

& Providing editing guidance & \\ 

 & \cellcolor{gray!30} Creating articles & \\

 & Writing codes for articles &  \\ \hline

\multirow{3}{*}{\centering Search}  &  \cellcolor{gray!30} Searching sources & \multirow{3}{*}{\parbox{9cm}{\textit{I think it’s better at searching for those than I am...it is a skill to phrase your queries in Google and know how to find things.} (P04)}} \\

& Suggesting images & \\

& \cellcolor{gray!30} Searching information & \\ \hline


\multirow{9}{*}{\centering Refine} & {Copyediting} & \textit{Google cannot do it. DuckDuckGo cannot do it. But language models are quite often able to do it well, and give me the short name of a thing.} (P07)  \\

& \cellcolor{gray!30} {Formatting} & \textit{In the old days, I would have copied this thing, put it into a plain text file, changed it to a CSV, uploaded it into R, and written a script. But with ChatGPT, I can just post the whole thing to ChatGPT, say I want it in this format.} (P04) \\

& {Translating} & \textit{You use wikitext, like square brackets to link to a page and you use single quotation marks to make something bold or italic and all this. Google Translate does not understand this wikitext language, but ChatGPT does.} (P05) \\

\bottomrule
\end{tabular}
\caption{Selected LLM Use Cases Reported by Participants. \textit{The distinction between generation and refinement lies in whether the content exists prior to interacting with the LLMs. In generation, LLMs are used to produce new text from scratch. In refinement, editors input existing content to receive feedback or suggestions for improvement. Copyediting includes grammatical improvements, rephrasing suggestions for repetitive words or longer description, brainstorming alternative expressions.}}
\label{tab:uses}
\end{table}

\subsection{\textbf{RQ1}: Contributing content}
Our findings for \textbf{RQ1} show that experience level mediates how LLM use influences content contribution. For experienced editors, LLMs extend cognitive capacities, enabling exploration of new topics and improving confidence and quality. In contrast, newcomers often rely on LLMs to fill in their knowledge gaps, as LLMs reduce the barriers to entry and provide a sense of guidance. 

\subsubsection{Experienced editors venture into new topics and tasks.}
LLMs demonstrated the ability to guide editors toward content they wouldn't have otherwise considered, by providing useful contextual information without requiring proactive efforts from the editors. P05 shared: ``\textit{The articles that I created using AI, I probably would not have created them}." This is especially true when available resources are limited for the exact tasks. P08, who was writing code for supplementary graphs to improve article quality, didn’t know anyone to consult about troubleshooting his R code, especially as it might need a long time to fix the bugs. LLMs served as a support tutor, as he reflected: 

\begin{quote}
``\textit{[ChatGPT] is basically the next closest thing to [people who can help]. It brings up different packages that I wouldn't consider...[LLM] definitely directed me to things that I would have never been able to think of on my own.}"  
\end{quote}

In this way, LLMs enabled him to engage in tasks he would not have attempted alone. P03 shared a similar anecdote. As he went down a rabbit hole on fruits and vegetables, LLMs suggested a new section he didn’t even know about and would not think of: 

\begin{quote}
``\textit{[ChatGPT] suggested a section where I didn't even know that a specific vegetable had been - a hybrid of this vegetable created another one that was not in the article…I looked to make sure the information was factually correct, and I put that in there.}" 
\end{quote}

For P03, he explored novel topics besides his typical contributions in business articles. The capacity of LLMs to introduce new, previously overlooked perspectives enhances editors’ creativity and encourages them to venture beyond their usual domains, most importantly, add the piece of knowledge in Wikipedia. 

\subsubsection{Experienced editors are exposed to new perspectives.} 
In addition to helping editors move beyond their typical domains, LLMs extend cognitive capacity by introducing unexpected and insightful perspectives. P08 recounted LLMs supporting him to reflect on his own text in a different way, for instance, by suggesting what is interesting about his text: 

\begin{quote}
``\textit{I’m asking LLMs for ideas...``What’s interesting in this big block of text I’ve written?" Honestly, I spent 48 hours staring at my computer screen writing this article. I’ve kind of bored myself to death with it.}"
\end{quote}

LLMs could also serve as a simulated average reader, offering pluralistic perspectives that editors find helpful in evaluating their content. P09 noted, ``\textit{I regularly ask LLMs...when I want a pluralistic opinion. You want to know what the average reader would think}." 

\subsubsection{Experienced editors are more confident in unfamiliar areas.}
This cognitive support, in turn, increased confidence and satisfication for experienced editors, even in domains they are less familiar with. P03 felt more capable of editing complex topics such as medical and legal content: ``\textit{I can digest the information if I don’t understand...and run through LLMs to simplify for myself}." P04 treats LLMs as an additional validation mechanism: ``\textit{LLMs are a good extra filter to make sure there’s not something you overlook or mistakes you are making}." P08 remembered editing without LLMs took endless iterations, as he would ``\textit{write, write, try to get through it}," especially when he got stuck. The presence of LLMs helped him feel satisfied with his work sooner, even improving his sleep routine: ``\textit{it allows me to feel happy with what I’ve written earlier...thus allow me to go to bed earlier}." 

\subsubsection{LLMs enhance contribution quality for experienced editors.} Multiple experienced participants observed that the quality of their contributions has improved. For example, P09 specified that his accuracy on topics where he lacked in-depth knowledge ``\textit{had definitely improved}." P03 agreed with P09, and shared that fewer post-edit corrections from other editors signaled quality improvement: ``\textit{My edits over the past two years - the quality has improved because fewer people have had to come behind me and clean up}." 

 

\subsubsection{LLMs lower entry barriers for newcomers.}
New editors themselves frequently shared that LLMs encouraged them to edit Wikipedia. For example, P14 described the challenges related to information access in his local context, and how LLMs helped with research that would otherwise take days: 

\begin{quote}
``\textit{LLMs encourages me to work on Wikipedia...I have to go to libraries to look for information...I must pay for the transportation and a lot of other things...Before I go to the library, I’ll check on AI to know more about the topic.}" (P14) 
\end{quote}

P01 found that proofreading using LLMs made editing easier, particularly given that English was not his native language: ``\textit{English is not my native language...I have to proofread it on ChatGPT}." More experienced editors agreed that LLMs lowered the barriers for entry, as P12, P09, P04 and P05 pointed out that LLMs ``\textit{increase the ability of new contributors to contribute, as it breaks down a lot of the barriers} (P12)."

\subsubsection{New editors rely on LLMs to fill in their knowledge gaps.}
As LLMs are able to lower the technical and linguistic barriers, newcomers tend to rely on LLMs to fill in their knowledge gaps, often in generating and searching for content. Editors observed that newcomers may use LLMs to generate content. P04 observed that many newcomers were using LLMs to draft articles: ``\textit{There’s an increased expansion...the vast majority of them just write a prompt like, hey, ChatGPT, can you write a Wikipedia article on this?}" Newcomers also leveraged LLMs to search for knowledge around the subject topics, and specific wiki knowledge. For instance, P10 shared that in order to ensure high quality contribution, he ``\textit{use LLMs as my research platform to know more about what I’m looking for}." P01 emphasized that he used ChatGPT to ``\textit{increase my knowledge around the subjects}." He described LLMs as a wise mentor he could learn from.: 

\begin{quote}
``\textit{I take the information and integrate it into my thoughts, like speaking with a wise person..I don’t have the luxury of speaking with super intellectuals around myself...I compensate for the gap of knowledge.}" (P01) 
\end{quote}

Similarly, P14 perceived LLMs to be a ``school" for him, specifying that LLMs allowed him to ``\textit{know more about anything inside my house}." He demonstrated this by sharing an example of him deciding to create an article on English Wikipedia, and asking LLMs ``\textit{how do I start doing this?}" and learned that the articles needed ``\textit{citations, notability and other things}." LLMs can be a guide for newcomers. P15 reasoned that ``\textit{I need a guide...If there is anything that could guide me for better content.}"

\subsection{\textbf{RQ2}: Conforming to community norms}

To answer \textbf{RQ2}, our findings showed that editors rely on core content policies and guidelines to navigate the unclear norms around LLM usage. They recognize that LLMs could violate these standards, indicating that LLM use is situated within the Wikipedia editing ecosystem. Participants described three strategies to make editorial judgments: evaluation, verification, and modification. As a result, LLMs challenged new editors, as they had not yet developed skills to make complex judgments and align to community standards. 

\subsubsection{LLMs fail to align with core Wikipedia content policies.} 
Participants consistently described limitations of LLMs from the lens of violating core Wikipedia policies, as P02 summarized, ``\textit{LLM output has many flaws and often violates Wikipedia policies}." The most frequently mentioned concerns were related to the Neutral Point of View (NPOV) \citep{npov}, Verifiability \citep{verifiability}, and No Original Research (NOR) \citep{nor}. 

NPOV \citep{npov} is defined as “representing all the significant views on a topic fairly and proportionately without editorial bias.” LLMs often produce language that violates NPOV due to an overly positive or promotional tone. P05 noticed that LLMs tend to use phrases such as ``\textit{it’s one of the best” or “there are so many possibilities}," which are discouraged unless supported by reliable citations. He added that such content often contains ``\textit{puffery peacock terms}," especially when asking LLMs to write something from scratch. P07 agreed with P05, as he observed that LLMs could use English expressions not typically found in Wikipedia. This aligns with prior work \citep{ashkinaze2024seeing}, which found that LLMs are limited in detecting and generating content that conforms to NPOV like a community expert. 

Verifiability \citep{verifiability} refers to the ability of a claim to be proven right or wrong, typically through the presence or availability of sources. It is compromised when LLMs cite unreliable sources or fabricated sources. P07 criticized a LLM named Perplexity. While sometimes it ``\textit{doesn’t give any sources}," other times when it gives sources, they would not meet the Wikipedia standards, for being ``\textit{unreliable junk commercial websites that are made not for providing correct or verified information, but just for gathering clicks and showing advertisements}."

LLMs are prone to hallucination, especially for obscure content. This violates NOR \citep{nor} which prohibits original research that finds no reliable or published sources to support. More importantly, LLMs' tendency to hallucinate contradicts the very core value of knowledge integrity. P04 specified that LLM-generated content was ``\textit{not necessarily accurate}" as LLMs ``\textit{can hallucinate and make up sources}." P02 and P03 also mentioned hallucination, as P02 explained ``\textit{when the model invents non-existing facts and invents non-existing references to support claims}." P08 shared specifically about LLMs struggling with less common programming languages, noting their limited performance beyond languages like Python or R. 

\subsubsection{Participants’ strategies to make editorial judgements.} 
To cope with the risk that LLMs may violate policies and guidelines, our participants emphasized their responsibility as gatekeepers. P03 stressed, “``\textit{even if you have all the tools available, you are still the last one to make the decision whether this should go in or not}." In practice, editors used three key strategies when working with LLM-generated content: evaluation, verification, and modification.

Evaluation involved assessing whether LLM suggestions improved the readability and coherence of the content. P08 put it as ``\textit{make judgement about whether the LLM content makes sense}." P02 developed a practice of ``\textit{evaluating whether any of the [suggestions] would improve the writing}." P09 evaluated the quality in terms of consistency as he ``\textit{regenerated the answers to see if it’s consistent, [and] compared the answers of multiple chatbots}." 

While evaluation ensures readability and coherence, verification checks the content’s factual accuracy and reliability. P09 and P04 both stressed ``\textit{verifying that everything makes sense and is supported by sources} (P09)." P07 noted the necessity to `\textit{check every word that LLM wrote because of verifiability}." He mentioned specific strategies for verification: reaching out to friends with questions, and cross-referencing terminologies or phrases via web searches.

Additionally, participants intensively modified the outputs before incorporating them into Wikipedia to align with Wikipedia’s tone and style. P16 noticed that LLMs tend to be overall more positive, and depicted modification as neutralizing the tone of the text. P07 would ``\textit{heavily edit}" the output by himself before publishing. P03, P04, and P05 would modify the text to better fit the context, such as ``\textit{wiki format} (P04)" the links and fixing typos in wikitext, as ``\textit{ChatGPT could translate the wikilinks or reference wrongly so I have to update and change them} (P05)."

\subsubsection{Newcomers as gatekeepers?}
Being gatekeepers for LLM-generated content requires expertise and wiki knowledge, specifically around Wikipedia norms and standards.  However, this role is high-stakes and unfamiliar to newcomers, as they have not yet developed these skills. For example, newcomers recognize that they did not know all the guidelines and principles as P01 pointed out ``\textit{I am still developing my skills. I’m new to Wikipedia, so I may not know all the guidelines and principles}." P03, an experienced editor, talked about how he made decisions when interacting with LLMs: ``\textit{I am the big decider of what I know in my gut what should be in Wikipedia}." And he pointed out that editors with many years of contributions would have the expertise to tell whether materials should be in Wikipedia or not. He then contrasted editors with different levels of expertise: ``\textit{I have 20 years of editing right now. So, the tool is much more of an enhancement to me because I know what's allowed and not allowed quickly. I can accept or reject a suggested edit...[but] a new user would go through and just accept, accept, accept the edit}." P02 mentioned that ``\textit{inexperienced editors often lack the ability to evaluate whether an output is appropriate...This creates problems when inexperienced editors unaware of the limitations of LLMs rely on them to make contributions without critical examination}." This suggests that newcomers lack the ability to critically evaluate and examine LLM-generated content and may over-rely on LLM outputs, which leads to challenges in producing high-quality content. This is well documented in an example P15, as a newcomer himself, shared: 

\begin{quote}
    ``\textit{I was trying to get good content out of it, but it had so many unwanted things as well. As I'm learning, I'm not that clear about the content exactly, what should I filter out. Sometimes even if I read it multiple times, I do not know whether it is needed or not, or whether the content is good or not, if this particular explanation is useful or not. Then after this editor pointed it out, I noticed that’s not what [Wikipedia] wanted.}" (P15)
\end{quote}

\subsection{\textbf{RQ3}: Community engagement}

To answer \textbf{RQ3}, we found that other editors’ response to LLM-assisted edits further revealed the participation divide between new and experienced editors. Other editors praise and accept LLM-assisted contributions from experienced editors, yet call out and reject those made by newcomers. The ability of conforming AI-generated content to standards leads to difference in other members' perception on whether the content is AI-generated, resulting in the differences in the response. Such dynamics may contribute to the overall sensitivity and confusion around LLM usage in the community.

\subsubsection{Other editors responded to newcomers with rejections.}
New editors faced heightened scrutiny when using LLMs, as they found themselves in a position where other editors quickly found out that they had used LLMs. P11 recounted his experience: ``\textit{As soon as I pushed the article out there as a draft, I got a message from somebody in the community saying, “oh, it looks like you created this with an LLM and we don't want that kind of material.}" He reflected, ``\textit{I'm glad that they're being careful about it, but it’s clear that they’re sensitive.}" Likewise, P01 was told his contribution was ``\textit{machine generated,}" and P10 was criticized for making ``\textit{promotional and essay-like articles}." P15 shared that another editor quickly identified his use of ChatGPT, which prompted him to remove all LLM-generated content. As P09 noted, ``\textit{most of the poorly written AI content is already removed...In fact, it gets removed pretty quickly}," suggesting that contributions flagged as LLM-generated are assumed to be low quality and are swiftly deleted. 

\subsubsection{Other editors responded to experienced editors with approval.}
In contrast, experienced editors often received positive feedback. P16 shared that one editor praised him for copyediting a section this editor had written, ``\textit{better than he had ever seen before}." He clarified: ``\textit{it was chatgpt that had done it}." P04 also shared his positive experience: ``\textit{In general, LLM-based edits I made are well received. They're not really received any differently than in my non-LLM based edits.}" P07 noted that no one noticed or identified that he had used LLMs, due to extensive post-editing. P09 received a similar response, as no one complained about his usage of LLMs, and his edits were rarely reverted. However, However, even experienced editors occasionally faced false accusations, as P03 noted: ``\textit{being called out wrongly infuriated me}." When the edits seemed to be generated by LLMs, even experienced editors were called out, underscoring the community's sensitivity to LLMs. 

\subsubsection{Overall, the community remains sensitive and confused about LLM uses.}
The general sensitivity and confusion around LLMs may stem from a lack of consensus on what it means to use LLMs for editing. P04 remarked that no one knows exactly what ``using LLMs" entails: ``\textit{it's so ambiguous...people might assume that [LLM assistance] means you copy and pasted text from an LLM prompt without any review or understanding}." The assumption can lead to mistrust, even when the content is factually accurate. P09 brought up an observation for editors attacking good edits because they were known to be generated by LLMs: ``\textit{I have once seen spiteful comments...other people severely criticized [the article] for being AI-generated}," and he concluded ``\textit{generally speaking, the vibe around LLMs is poor among other Wikipedia editors}." This suggests that the community cares not only the quality of the content, but also about its origin.

\section{Discussion}

\subsection{The paradox of participation}
Our findings reveal a paradox of participation: LLMs simultaneously lower barriers to entry while increasing the demands of contributing, especially for newcomers who already struggle to engage with the community \citep{halfaker2013rise, morgan2018evaluating, steinmacher2015systematic}. We discuss this paradox of participation in three interrelated elements:
\begin{itemize}
\item {LLMs interrupt traditional learning pathways for newcomers that support gradual skill acquisition for newcomers. }
\item {LLMs shift the focus from peripheral tasks to editorial judgment, requiring newcomers to make normative decisions before developing core competencies. } 
\item {As a result, LLMs exacerbate a participation divide, enabling experienced editors to thrive while marginalizing newcomers. }
\end{itemize}
These dynamics challenge expectations of Legitimate peripheral participation (LPP) \citep{lave1991situated, bryant2005becoming, halfaker2013making} and situated learning \citep{lave1991situated, roberts2014community}. In addition, they reveal nuanced aspects of the second-level digital divide \citep{hargittai2001second} in socio-technical systems in peer production platforms. 

\subsubsection{Interrupted learning pathways for newcomers (Situated learning)}
Before the widespread use of LLMs, learning in Wikipedia for newcomers followed a gradual and scaffolded path. Legitimate Peripheral Participation (LPP) \citep{lave1991situated, bryant2005becoming} states that newcomers start from low-risk and small tasks, and progressively take on more responsibilities. Built on LPP, situated learning \citep{lave1991situated, roberts2014community} emphasizes that learning is social: individuals gain skills and knowledge through interactions with other members in collaborative work. 

In Wikipedia, newcomers typically begin with low-stakes tasks such as fixing errors and improve grammars \citep{bryant2005becoming}. Through receiving feedback from other community members, the gradual learning pathway to participation enables newcomers to gain technical skills, understand community norms, and collaborate with others. Given Wikipedia’s complex norms and standards, this learning pathway is especially important for newcomers to develop normative understandings of policies and guidelines and gain legitimacy \citep{halfaker2013making, preece2009reader}. 

However, LLMs disrupt this learning trajectory by enabling new editors to directly contribute to complex tasks. For new editors, LLMs offer the promise of access to knowledge. For instance, our results show that newcomers have increased access to linguistic support (especially for non-native speakers), contextual suggestions (wikitext), and source search. These affordances enable them to overcome traditional entry barriers, increase confidence, and thus take on complex tasks such as drafting new articles or synthesizing sources for their initial stages of contributions. As a result, rather than working toward the more complicated tasks step by step, newcomers now can directly make these contributions. 

In this way, LLMs accelerate participation. However, they simultaneously shortcut the social situated learning that Wikipedia had relied on to support newcomers’ participation. LLMs help users perform harder tasks, but they skip the essential steps to teach users on how to understand Wikipedia norms, and interpret content policies and guidelines. In exchange for immediate access to support, newcomers miss opportunities to internalize Wikipedia as a social and epistemic community. 

\subsubsection{A shift from peripheral tasks to editorial judgment (LPP)}
While LLMs enable newcomers to bypass low-risk contributions, they impose new demands, which fundamentally shift the nature of participation. Instead of easing the learning curve, LLMs thrust newcomers into an unexpectedly high-stakes and unfamiliar role that requires editorial judgment. Editorial judgment demands deeper understanding of Wikipedia’s social norms, values and culture manifested in policies and guidelines as gatekeepers \citep{bryant2005becoming}. To meet community standards, newcomers are now responsible for verifying, evaluating, and modifying AI-generated content before publishing it. These responsibilities assume both epistemic and social maturity. 

Previous AI tools on Wikipedia, such as SuggestBot \citep{cosley2007suggestbot}, ORES \citep{halfaker2020ores}, Vandal Fighter \citep{tsvetkova2017even}, inherently align and reinforce community norms. In contrast, LLMs delegate the burden of judgment to editors. These judgments are higher-stake and central to community culture, which are developed after gradual participation. Our findings show that LLMs interrupt learning pathways for newcomers, which potentially results in further lack of competencies in making such judgments. Yet, newcomers are challenged to make these normative decisions from the outset, with little feedback and guidance. As a result, LLMs not only make it harder for newcomers to participate, but also fundamentally change what participation entails, which challenges our assumptions about participation in communities of practice.

\subsubsection{A new participation divide mediated by expertise (Second-level digital divide)}
While new editors face challenges for their LLM-assisted contributions, more experienced editors are able to benefit from LLMs. Our findings suggest that experienced contributors expand their participation through new topics and overcome their writer’s blocks. We articulate this difference to originate from the level of expertise. More experienced editors are familiar with Wikipedia policies and guidelines to critically evaluate AI output and adapt AI output to Wikipedia standards. Therefore, their participation is enhanced. 

This participation divide stems not from access, but from unequal ability to effectively use LLMs. The lack of expertise to use LLMs responsibly and acceptable to Wikipedia seem to be at the root of the problem. This mirrors the concepts of the second-level digital divide \citep{hargittai2001second}, which specifies the disparity between individuals with different skills and knowledge to utilize technologies to fully benefit from them, even if the access remains same. As a result, LLMs widen existing gaps between newcomers and experienced editors in participating in the community from the ability to manage AI outputs. 

\subsection{Design implications}
The paradox is both interesting and novel: LLMs empower newcomers to do more, but also demand more of them. LLMs enable newcomers to participate, but fail to support legitimate participation. Our results underscore a key insight: increased access does not guarantee better participation. In communities of practice like Wikipedia, expertise mediates the relationship between access to meaningful contributions. At the same time, LLMs demonstrate potential for positive uses. For instance, they lower technical barriers, encourage contributions, and enhance participation across domains. Thus, to account for both the potential and challenges of leveraging LLMs, we propose several design implications. 

\subsubsection{Scaffold participation through incremental guidance}
Good-faith newcomers desire to learn and participate in communities, but when LLMs flatten complex editorial processes into generated outputs from one simple prompt, they remove crucial opportunities for learning. Rather than doing the work for newcomers, LLMs should scaffold the process to guide them, given that scaffolding can lead to improved writing quality \citep{dhillon2024shaping}. 

For example, if an editor asks LLMs to generate a Wikipedia article, the system could decompose the task into smaller steps, e.g., finding sources, summarizing content, and structuring the article. Instead of directly returning a complete draft, the LLM could prompt: “Which sources would you consider for this topic?” In doing so, the LLMs act as facilitators, enabling newcomers to learn through their conversational interactions. 

\subsubsection{Teach community norms through interactions}
In addition to guiding editors to do the work, LLMs should help them understand how to do it right. This means embedding normative feedback into the interactions. When newcomers rely on LLMs, the system can serve as a reflective layer that encourages them to evaluate whether their contributions align with community standards. 

There are several ways LLMs can provide such feedback. LLMs can highlight problematic portions to draw editors’ attention, provide both acceptable and unacceptable examples, and ask guided questions that allow editors to reflect by themselves. In these ways, LLMs support norm understanding and adaptation, which is a vital element for communities of practice. 

\subsubsection{Be aware of and personalize based on user expertise}
Finally, context-aware LLMs should account for who is asking. Expertise matters, especially in online communities. While experienced editors may benefit from direct outputs, newcomers might need more guardrails. LLMs should adapt their responses accordingly. 

For instance, LLMs can refrain from directly generating content for newcomers, but instead, walking through each step with newcomers, thus allowing them to learn by doing the actual work. In contrast, LLMs may be a good co-writer for an experienced editor, with higher degree of freedom in what LLMs produce. Such systems do not aim to restrict newcomers, but to ensure responsible participation for newcomers. 

We outlined three design implications for future LLM-based assistants in knowledge production communities, which aligned with WikiMedia Foundation's strategy regarding AI tools for editors \citep{wikimedia}, especially in terms of helping newcomers understand policies and provide feedback to their edits. Our implications respect and are grounded in the traditional learning pathways stated by Legitimate Peripheral Participation and Situated Learning. Accordingly, we recommend designing such assistants to guide newcomers through the incremental trajectory from peripheral tasks to more central and complex contributions. However, our results seem to hint at a new reality where newcomers bypass peripheral tasks altogether. Newcomers wanted to create articles and LLMs may just be a tool they utilize to achieve their goals. If not LLMs, they would still start contributing complex tasks from day one. This leaves us to ponder, if this bypassing behavior is true, how should the community and researchers respond? We leave future research to explore this emerging dynamic. 

\subsection{Limitations and future work}
Our study provides an in-depth understanding of how editors interact with LLMs in their knowledge production on Wikipedia. However, we acknowledge two limitations for our study. 

First, while our qualitative approach enabled us to understand nuanced editor experiences and perceptions, it did not quantify the prevalence of observed phenomena, which we recognized to be an important question to uncover. However, our categories and themes for use cases could serve as a foundation for future research. Future research could utilize our use cases (see Table ~\ref{tab:uses} for LLM use cases) to develop surveys to assess the generalizability and prevalence of these use cases and interaction patterns. 

Second, our study only focused on editors who have used LLMs in their editing process. This meets our expectations and addresses our needs as we aim to understand user experience. However, this might introduce bias towards the confidence level of technology adoptions, and most importantly underrepresented editors who deliberately choose to not engage with LLMs. Future studies should include perspective from these editors, especially those who initially chose to use LLMs but later stopped.

\section{Conclusion}
To address the gap in understanding the impact of adopting generative AI on knowledge contributors’ participation in communities of practice, we recruited 16 participants who had used LLMs in their editing process from Wikipedia and conducted semi-structured interviews with them. We asked about their perception, adoption, and interaction with LLMs, and other editors’ response to their LLM-assisted edits. We found that LLMs introduced a participation divide between new and experienced editors manifested in contributing content, enforcing norms, and other editors' engagement. For newcomers, the paradox of participation indicated that LLMs 1) interrupt their learning pathway, and 2) shift traditional peripheral tasks into central tasks requiring editorial judgment. This challenge further escalates as experienced editors are able to enhance their participation by 1) exploring new topics, 2) gaining multiple perspectives and 3) increasing confidence. We offered design implications to mitigate the participation gap, including scaffolding complex tasks for newcomers during interaction, educating community norms and standards, and considering expertise as an important part of context. Our study demonstrates the importance of user experience research in shaping equitable AI integration in communities of practice, and highlights future opportunities for designing LLM-powered tools that not only support production but also foster community collaboration.

\bibliographystyle{ACM-Reference-Format}
\bibliography{sample-base}


\begin{thebibliography}{98}


\ifx \showCODEN    \undefined \def \showCODEN     #1{\unskip}     \fi
\ifx \showISBNx    \undefined \def \showISBNx     #1{\unskip}     \fi
\ifx \showISBNxiii \undefined \def \showISBNxiii  #1{\unskip}     \fi
\ifx \showISSN     \undefined \def \showISSN      #1{\unskip}     \fi
\ifx \showLCCN     \undefined \def \showLCCN      #1{\unskip}     \fi
\ifx \shownote     \undefined \def \shownote      #1{#1}          \fi
\ifx \showarticletitle \undefined \def \showarticletitle #1{#1}   \fi
\ifx \showURL      \undefined \def \showURL       {\relax}        \fi
\providecommand\bibfield[2]{#2}
\providecommand\bibinfo[2]{#2}
\providecommand\natexlab[1]{#1}
\providecommand\showeprint[2][]{arXiv:#2}

\bibitem[Agarwal et~al\mbox{.}(2024)]%
        {agarwal2024litllms}
\bibfield{author}{\bibinfo{person}{Shubham Agarwal}, \bibinfo{person}{Gaurav Sahu}, \bibinfo{person}{Abhay Puri}, \bibinfo{person}{Issam~H Laradji}, \bibinfo{person}{Krishnamurthy~Dj Dvijotham}, \bibinfo{person}{Jason Stanley}, \bibinfo{person}{Laurent Charlin}, {and} \bibinfo{person}{Christopher Pal}.} \bibinfo{year}{2024}\natexlab{}.
\newblock \showarticletitle{LitLLMs, LLMs for Literature Review: Are we there yet?}
\newblock \bibinfo{journal}{\emph{Transactions on Machine Learning Research}} (\bibinfo{year}{2024}).
\newblock


\bibitem[Alavi and Leidner(2001)]%
        {alavi2001knowledge}
\bibfield{author}{\bibinfo{person}{Maryam Alavi} {and} \bibinfo{person}{Dorothy~E Leidner}.} \bibinfo{year}{2001}\natexlab{}.
\newblock \showarticletitle{Knowledge management and knowledge management systems: Conceptual foundations and research issues}.
\newblock \bibinfo{journal}{\emph{MIS quarterly}} (\bibinfo{year}{2001}), \bibinfo{pages}{107--136}.
\newblock


\bibitem[Ashkinaze et~al\mbox{.}(2024)]%
        {ashkinaze2024seeing}
\bibfield{author}{\bibinfo{person}{Joshua Ashkinaze}, \bibinfo{person}{Ruijia Guan}, \bibinfo{person}{Laura Kurek}, \bibinfo{person}{Eytan Adar}, \bibinfo{person}{Ceren Budak}, {and} \bibinfo{person}{Eric Gilbert}.} \bibinfo{year}{2024}\natexlab{}.
\newblock \showarticletitle{Seeing like an ai: How llms apply (and misapply) wikipedia neutrality norms}.
\newblock \bibinfo{journal}{\emph{arXiv preprint arXiv:2407.04183}} (\bibinfo{year}{2024}).
\newblock


\bibitem[Aubin Le~Qu{\'e}r{\'e} et~al\mbox{.}(2024)]%
        {aubin2024llms}
\bibfield{author}{\bibinfo{person}{Marianne Aubin Le~Qu{\'e}r{\'e}}, \bibinfo{person}{Hope Schroeder}, \bibinfo{person}{Casey Randazzo}, \bibinfo{person}{Jie Gao}, \bibinfo{person}{Ziv Epstein}, \bibinfo{person}{Simon~Tangi Perrault}, \bibinfo{person}{David Mimno}, \bibinfo{person}{Louise Barkhuus}, {and} \bibinfo{person}{Hanlin Li}.} \bibinfo{year}{2024}\natexlab{}.
\newblock \showarticletitle{LLMs as research tools: Applications and evaluations in HCI data work}. In \bibinfo{booktitle}{\emph{Extended Abstracts of the CHI Conference on Human Factors in Computing Systems}}. \bibinfo{pages}{1--7}.
\newblock


\bibitem[Ayers et~al\mbox{.}(2008)]%
        {ayers2008wikipedia}
\bibfield{author}{\bibinfo{person}{Phoebe Ayers}, \bibinfo{person}{Charles Matthews}, {and} \bibinfo{person}{Ben Yates}.} \bibinfo{year}{2008}\natexlab{}.
\newblock \bibinfo{booktitle}{\emph{How Wikipedia works: And how you can be a part of it}}.
\newblock \bibinfo{publisher}{No Starch Press}.
\newblock


\bibitem[Beschastnikh et~al\mbox{.}(2008)]%
        {beschastnikh2008wikipedian}
\bibfield{author}{\bibinfo{person}{Ivan Beschastnikh}, \bibinfo{person}{Travis Kriplean}, {and} \bibinfo{person}{David McDonald}.} \bibinfo{year}{2008}\natexlab{}.
\newblock \showarticletitle{Wikipedian self-governance in action: Motivating the policy lens}. In \bibinfo{booktitle}{\emph{Proceedings of the International AAAI Conference on Web and Social Media}}, Vol.~\bibinfo{volume}{2}. \bibinfo{pages}{27--35}.
\newblock


\bibitem[Biermann et~al\mbox{.}(2022)]%
        {biermann2022tool}
\bibfield{author}{\bibinfo{person}{Oloff~C Biermann}, \bibinfo{person}{Ning~F Ma}, {and} \bibinfo{person}{Dongwook Yoon}.} \bibinfo{year}{2022}\natexlab{}.
\newblock \showarticletitle{From tool to companion: Storywriters want AI writers to respect their personal values and writing strategies}. In \bibinfo{booktitle}{\emph{Proceedings of the 2022 ACM Designing Interactive Systems Conference}}. \bibinfo{pages}{1209--1227}.
\newblock


\bibitem[Brachman et~al\mbox{.}(2024)]%
        {brachman2024knowledge}
\bibfield{author}{\bibinfo{person}{Michelle Brachman}, \bibinfo{person}{Amina El-Ashry}, \bibinfo{person}{Casey Dugan}, {and} \bibinfo{person}{Werner Geyer}.} \bibinfo{year}{2024}\natexlab{}.
\newblock \showarticletitle{How knowledge workers use and want to use LLMs in an enterprise context}. In \bibinfo{booktitle}{\emph{Extended Abstracts of the CHI Conference on Human Factors in Computing Systems}}. \bibinfo{pages}{1--8}.
\newblock


\bibitem[Bronner et~al\mbox{.}(2012)]%
        {bronner2012cosyne}
\bibfield{author}{\bibinfo{person}{Amit Bronner}, \bibinfo{person}{Matteo Negri}, \bibinfo{person}{Yashar Mehdad}, \bibinfo{person}{Angela Fahrni}, {and} \bibinfo{person}{Christof Monz}.} \bibinfo{year}{2012}\natexlab{}.
\newblock \showarticletitle{Cosyne: Synchronizing multilingual wiki content}. In \bibinfo{booktitle}{\emph{Proceedings of the Eighth Annual International Symposium on Wikis and Open Collaboration}}. \bibinfo{pages}{1--4}.
\newblock


\bibitem[Brooks et~al\mbox{.}(2024)]%
        {brooks2024rise}
\bibfield{author}{\bibinfo{person}{Creston Brooks}, \bibinfo{person}{Samuel Eggert}, {and} \bibinfo{person}{Denis Peskoff}.} \bibinfo{year}{2024}\natexlab{}.
\newblock \showarticletitle{The Rise of AI-Generated Content in Wikipedia}.
\newblock \bibinfo{journal}{\emph{arXiv preprint arXiv:2410.08044}} (\bibinfo{year}{2024}).
\newblock


\bibitem[Bryant et~al\mbox{.}(2005)]%
        {bryant2005becoming}
\bibfield{author}{\bibinfo{person}{Susan~L Bryant}, \bibinfo{person}{Andrea Forte}, {and} \bibinfo{person}{Amy Bruckman}.} \bibinfo{year}{2005}\natexlab{}.
\newblock \showarticletitle{Becoming Wikipedian: transformation of participation in a collaborative online encyclopedia}. In \bibinfo{booktitle}{\emph{Proceedings of the 2005 ACM international conference on supporting group work}}. \bibinfo{pages}{1--10}.
\newblock


\bibitem[Butler et~al\mbox{.}(2008)]%
        {butler2008don}
\bibfield{author}{\bibinfo{person}{Brian Butler}, \bibinfo{person}{Elisabeth Joyce}, {and} \bibinfo{person}{Jacqueline Pike}.} \bibinfo{year}{2008}\natexlab{}.
\newblock \showarticletitle{Don't look now, but we've created a bureaucracy: the nature and roles of policies and rules in wikipedia}. In \bibinfo{booktitle}{\emph{Proceedings of the SIGCHI conference on human factors in computing systems}}. \bibinfo{pages}{1101--1110}.
\newblock


\bibitem[Caine(2016)]%
        {caine2016local}
\bibfield{author}{\bibinfo{person}{Kelly Caine}.} \bibinfo{year}{2016}\natexlab{}.
\newblock \showarticletitle{Local standards for sample size at CHI}. In \bibinfo{booktitle}{\emph{Proceedings of the 2016 CHI conference on human factors in computing systems}}. \bibinfo{pages}{981--992}.
\newblock


\bibitem[Chakrabarty et~al\mbox{.}(2022)]%
        {chakrabarty2022help}
\bibfield{author}{\bibinfo{person}{Tuhin Chakrabarty}, \bibinfo{person}{Vishakh Padmakumar}, {and} \bibinfo{person}{He He}.} \bibinfo{year}{2022}\natexlab{}.
\newblock \showarticletitle{Help me write a poem: Instruction tuning as a vehicle for collaborative poetry writing}.
\newblock \bibinfo{journal}{\emph{arXiv preprint arXiv:2210.13669}} (\bibinfo{year}{2022}).
\newblock


\bibitem[Chung et~al\mbox{.}(2022)]%
        {chung2022talebrush}
\bibfield{author}{\bibinfo{person}{John Joon~Young Chung}, \bibinfo{person}{Wooseok Kim}, \bibinfo{person}{Kang~Min Yoo}, \bibinfo{person}{Hwaran Lee}, \bibinfo{person}{Eytan Adar}, {and} \bibinfo{person}{Minsuk Chang}.} \bibinfo{year}{2022}\natexlab{}.
\newblock \showarticletitle{TaleBrush: Sketching stories with generative pretrained language models}. In \bibinfo{booktitle}{\emph{Proceedings of the 2022 CHI Conference on Human Factors in Computing Systems}}. \bibinfo{pages}{1--19}.
\newblock


\bibitem[Clarke and Braun(2017)]%
        {clarke2017thematic}
\bibfield{author}{\bibinfo{person}{Victoria Clarke} {and} \bibinfo{person}{Virginia Braun}.} \bibinfo{year}{2017}\natexlab{}.
\newblock \showarticletitle{Thematic analysis}.
\newblock \bibinfo{journal}{\emph{The journal of positive psychology}} \bibinfo{volume}{12}, \bibinfo{number}{3} (\bibinfo{year}{2017}), \bibinfo{pages}{297--298}.
\newblock


\bibitem[Cl{\'e}ment and Guitton(2015)]%
        {clement2015interacting}
\bibfield{author}{\bibinfo{person}{Maxime Cl{\'e}ment} {and} \bibinfo{person}{Matthieu~J Guitton}.} \bibinfo{year}{2015}\natexlab{}.
\newblock \showarticletitle{Interacting with bots online: Users’ reactions to actions of automated programs in Wikipedia}.
\newblock \bibinfo{journal}{\emph{Computers in Human Behavior}}  \bibinfo{volume}{50} (\bibinfo{year}{2015}), \bibinfo{pages}{66--75}.
\newblock


\bibitem[Cosley et~al\mbox{.}(2007)]%
        {cosley2007suggestbot}
\bibfield{author}{\bibinfo{person}{Dan Cosley}, \bibinfo{person}{Dan Frankowski}, \bibinfo{person}{Loren Terveen}, {and} \bibinfo{person}{John Riedl}.} \bibinfo{year}{2007}\natexlab{}.
\newblock \showarticletitle{SuggestBot: using intelligent task routing to help people find work in wikipedia}. In \bibinfo{booktitle}{\emph{Proceedings of the 12th international conference on Intelligent user interfaces}}. \bibinfo{pages}{32--41}.
\newblock


\bibitem[Daxenberger and Gurevych(2012)]%
        {daxenberger2012corpus}
\bibfield{author}{\bibinfo{person}{Johannes Daxenberger} {and} \bibinfo{person}{Iryna Gurevych}.} \bibinfo{year}{2012}\natexlab{}.
\newblock \showarticletitle{A corpus-based study of edit categories in featured and non-featured Wikipedia articles}. In \bibinfo{booktitle}{\emph{Proceedings of COLING 2012}}. \bibinfo{pages}{711--726}.
\newblock


\bibitem[Daxenberger and Gurevych(2013)]%
        {daxenberger2013automatically}
\bibfield{author}{\bibinfo{person}{Johannes Daxenberger} {and} \bibinfo{person}{Iryna Gurevych}.} \bibinfo{year}{2013}\natexlab{}.
\newblock \showarticletitle{Automatically classifying edit categories in Wikipedia revisions}. In \bibinfo{booktitle}{\emph{Proceedings of the 2013 Conference on Empirical Methods in Natural Language Processing}}. \bibinfo{pages}{578--589}.
\newblock


\bibitem[Dhillon et~al\mbox{.}(2024)]%
        {dhillon2024shaping}
\bibfield{author}{\bibinfo{person}{Paramveer~S Dhillon}, \bibinfo{person}{Somayeh Molaei}, \bibinfo{person}{Jiaqi Li}, \bibinfo{person}{Maximilian Golub}, \bibinfo{person}{Shaochun Zheng}, {and} \bibinfo{person}{Lionel~Peter Robert}.} \bibinfo{year}{2024}\natexlab{}.
\newblock \showarticletitle{Shaping human-ai collaboration: varied scaffolding levels in co-writing with language models}. In \bibinfo{booktitle}{\emph{Proceedings of the 2024 CHI Conference on Human Factors in Computing Systems}}. \bibinfo{pages}{1--18}.
\newblock


\bibitem[Dillahunt and Malone(2015)]%
        {dillahunt2015promise}
\bibfield{author}{\bibinfo{person}{Tawanna~R Dillahunt} {and} \bibinfo{person}{Amelia~R Malone}.} \bibinfo{year}{2015}\natexlab{}.
\newblock \showarticletitle{The promise of the sharing economy among disadvantaged communities}. In \bibinfo{booktitle}{\emph{Proceedings of the 33rd annual ACM conference on human factors in computing systems}}. \bibinfo{pages}{2285--2294}.
\newblock


\bibitem[Erickson and Kellogg(2000)]%
        {erickson2000social}
\bibfield{author}{\bibinfo{person}{Thomas Erickson} {and} \bibinfo{person}{Wendy~A Kellogg}.} \bibinfo{year}{2000}\natexlab{}.
\newblock \showarticletitle{Social translucence: an approach to designing systems that support social processes}.
\newblock \bibinfo{journal}{\emph{ACM transactions on computer-human interaction (TOCHI)}} \bibinfo{volume}{7}, \bibinfo{number}{1} (\bibinfo{year}{2000}), \bibinfo{pages}{59--83}.
\newblock


\bibitem[Forte et~al\mbox{.}(2012)]%
        {forte2012coordination}
\bibfield{author}{\bibinfo{person}{Andrea Forte}, \bibinfo{person}{Niki Kittur}, \bibinfo{person}{Vanessa Larco}, \bibinfo{person}{Haiyi Zhu}, \bibinfo{person}{Amy Bruckman}, {and} \bibinfo{person}{Robert~E Kraut}.} \bibinfo{year}{2012}\natexlab{}.
\newblock \showarticletitle{Coordination and beyond: social functions of groups in open content production}. In \bibinfo{booktitle}{\emph{Proceedings of the ACM 2012 conference on Computer Supported Cooperative Work}}. \bibinfo{pages}{417--426}.
\newblock


\bibitem[Geiger(2009)]%
        {geiger2009social}
\bibfield{author}{\bibinfo{person}{R~Stuart Geiger}.} \bibinfo{year}{2009}\natexlab{}.
\newblock \showarticletitle{The social roles of bots and assisted editing programs}. In \bibinfo{booktitle}{\emph{Proceedings of the 5th International Symposium on Wikis and Open Collaboration}}. \bibinfo{pages}{1--2}.
\newblock


\bibitem[Geiger(2018)]%
        {geiger2018lives}
\bibfield{author}{\bibinfo{person}{R~Stuart Geiger}.} \bibinfo{year}{2018}\natexlab{}.
\newblock \showarticletitle{The lives of bots}.
\newblock \bibinfo{journal}{\emph{arXiv preprint arXiv:1810.09590}} (\bibinfo{year}{2018}).
\newblock


\bibitem[Geiger and Halfaker(2013)]%
        {geiger2013levee}
\bibfield{author}{\bibinfo{person}{R~Stuart Geiger} {and} \bibinfo{person}{Aaron Halfaker}.} \bibinfo{year}{2013}\natexlab{}.
\newblock \showarticletitle{When the levee breaks: without bots, what happens to Wikipedia's quality control processes?}. In \bibinfo{booktitle}{\emph{Proceedings of the 9th International Symposium on Open Collaboration}}. \bibinfo{pages}{1--6}.
\newblock


\bibitem[Geiger and Halfaker(2017)]%
        {geiger2017operationalizing}
\bibfield{author}{\bibinfo{person}{R~Stuart Geiger} {and} \bibinfo{person}{Aaron Halfaker}.} \bibinfo{year}{2017}\natexlab{}.
\newblock \showarticletitle{Operationalizing conflict and cooperation between automated software agents in wikipedia: A replication and expansion of'even good bots fight'}.
\newblock \bibinfo{journal}{\emph{Proceedings of the ACM on human-computer interaction}} \bibinfo{volume}{1}, \bibinfo{number}{CSCW} (\bibinfo{year}{2017}), \bibinfo{pages}{1--33}.
\newblock


\bibitem[Geiger and Ribes(2010)]%
        {geiger2010work}
\bibfield{author}{\bibinfo{person}{R~Stuart Geiger} {and} \bibinfo{person}{David Ribes}.} \bibinfo{year}{2010}\natexlab{}.
\newblock \showarticletitle{The work of sustaining order in Wikipedia: The banning of a vandal}. In \bibinfo{booktitle}{\emph{Proceedings of the 2010 ACM conference on Computer supported cooperative work}}. \bibinfo{pages}{117--126}.
\newblock


\bibitem[Gero and Chilton(2019)]%
        {gero2019metaphoria}
\bibfield{author}{\bibinfo{person}{Katy~Ilonka Gero} {and} \bibinfo{person}{Lydia~B Chilton}.} \bibinfo{year}{2019}\natexlab{}.
\newblock \showarticletitle{Metaphoria: An algorithmic companion for metaphor creation}. In \bibinfo{booktitle}{\emph{Proceedings of the 2019 CHI conference on human factors in computing systems}}. \bibinfo{pages}{1--12}.
\newblock


\bibitem[Gero et~al\mbox{.}(2022)]%
        {gero2022sparks}
\bibfield{author}{\bibinfo{person}{Katy~Ilonka Gero}, \bibinfo{person}{Vivian Liu}, {and} \bibinfo{person}{Lydia Chilton}.} \bibinfo{year}{2022}\natexlab{}.
\newblock \showarticletitle{Sparks: Inspiration for science writing using language models}. In \bibinfo{booktitle}{\emph{Proceedings of the 2022 ACM Designing Interactive Systems Conference}}. \bibinfo{pages}{1002--1019}.
\newblock


\bibitem[Hale(2014)]%
        {hale2014multilinguals}
\bibfield{author}{\bibinfo{person}{Scott~A Hale}.} \bibinfo{year}{2014}\natexlab{}.
\newblock \showarticletitle{Multilinguals and Wikipedia editing}. In \bibinfo{booktitle}{\emph{Proceedings of the 2014 ACM conference on Web science}}. \bibinfo{pages}{99--108}.
\newblock


\bibitem[Halfaker and Geiger(2020)]%
        {halfaker2020ores}
\bibfield{author}{\bibinfo{person}{Aaron Halfaker} {and} \bibinfo{person}{R~Stuart Geiger}.} \bibinfo{year}{2020}\natexlab{}.
\newblock \showarticletitle{Ores: Lowering barriers with participatory machine learning in wikipedia}.
\newblock \bibinfo{journal}{\emph{Proceedings of the ACM on Human-Computer Interaction}} \bibinfo{volume}{4}, \bibinfo{number}{CSCW2} (\bibinfo{year}{2020}), \bibinfo{pages}{1--37}.
\newblock


\bibitem[Halfaker et~al\mbox{.}(2013a)]%
        {halfaker2013rise}
\bibfield{author}{\bibinfo{person}{Aaron Halfaker}, \bibinfo{person}{R~Stuart Geiger}, \bibinfo{person}{Jonathan~T Morgan}, {and} \bibinfo{person}{John Riedl}.} \bibinfo{year}{2013}\natexlab{a}.
\newblock \showarticletitle{The rise and decline of an open collaboration system: How Wikipedia’s reaction to popularity is causing its decline}.
\newblock \bibinfo{journal}{\emph{American behavioral scientist}} \bibinfo{volume}{57}, \bibinfo{number}{5} (\bibinfo{year}{2013}), \bibinfo{pages}{664--688}.
\newblock


\bibitem[Halfaker et~al\mbox{.}(2013b)]%
        {halfaker2013making}
\bibfield{author}{\bibinfo{person}{Aaron Halfaker}, \bibinfo{person}{Os Keyes}, {and} \bibinfo{person}{Dario Taraborelli}.} \bibinfo{year}{2013}\natexlab{b}.
\newblock \showarticletitle{Making peripheral participation legitimate: reader engagement experiments in wikipedia}. In \bibinfo{booktitle}{\emph{Proceedings of the 2013 conference on Computer supported cooperative work}}. \bibinfo{pages}{849--860}.
\newblock


\bibitem[Halfaker and Riedl(2012)]%
        {halfaker2012bots}
\bibfield{author}{\bibinfo{person}{Aaron Halfaker} {and} \bibinfo{person}{John Riedl}.} \bibinfo{year}{2012}\natexlab{}.
\newblock \showarticletitle{Bots and cyborgs: Wikipedia's immune system}.
\newblock \bibinfo{journal}{\emph{Computer}} \bibinfo{volume}{45}, \bibinfo{number}{03} (\bibinfo{year}{2012}), \bibinfo{pages}{79--82}.
\newblock


\bibitem[Hargittai(2001)]%
        {hargittai2001second}
\bibfield{author}{\bibinfo{person}{Eszter Hargittai}.} \bibinfo{year}{2001}\natexlab{}.
\newblock \showarticletitle{Second-level digital divide: Mapping differences in people's online skills}.
\newblock \bibinfo{journal}{\emph{arXiv preprint cs/0109068}} (\bibinfo{year}{2001}).
\newblock


\bibitem[Heather~Ford and Rizoiu(2023)]%
        {ford2023chatgpt}
\bibfield{author}{\bibinfo{person}{Michael~Davis Heather~Ford} {and} \bibinfo{person}{Marian-Andrei Rizoiu}.} \bibinfo{year}{2023}\natexlab{}.
\newblock \bibinfo{title}{Implications of ChatGPT for Knowledge Integrity on Wikipedia}.
\newblock
\urldef\tempurl%
\url{https://meta.wikimedia.org/wiki/Research:Implications_of_ChatGPT_for_knowledge_integrity_on_Wikipedia}
\showURL{%
\tempurl}
\newblock
\shownote{Accessed: 2025-05-01}.


\bibitem[Hsieh et~al\mbox{.}(2023)]%
        {hsieh2023nip}
\bibfield{author}{\bibinfo{person}{Jane Hsieh}, \bibinfo{person}{Joselyn Kim}, \bibinfo{person}{Laura Dabbish}, {and} \bibinfo{person}{Haiyi Zhu}.} \bibinfo{year}{2023}\natexlab{}.
\newblock \showarticletitle{" Nip it in the Bud": Moderation Strategies in Open Source Software Projects and the Role of Bots}.
\newblock \bibinfo{journal}{\emph{Proceedings of the ACM on Human-Computer Interaction}} \bibinfo{volume}{7}, \bibinfo{number}{CSCW2} (\bibinfo{year}{2023}), \bibinfo{pages}{1--29}.
\newblock


\bibitem[Jin et~al\mbox{.}(2025)]%
        {jin2025high}
\bibfield{author}{\bibinfo{person}{Fangzhou Jin}, \bibinfo{person}{Lanfang Sun}, \bibinfo{person}{Yunqiu Pan}, {and} \bibinfo{person}{Chin-Hsi Lin}.} \bibinfo{year}{2025}\natexlab{}.
\newblock \showarticletitle{High Heels, Compass, Spider-Man, or Drug? Metaphor Analysis of Generative Artificial Intelligence in Academic Writing}.
\newblock \bibinfo{journal}{\emph{Computers \& Education}} (\bibinfo{year}{2025}), \bibinfo{pages}{105248}.
\newblock


\bibitem[Kiene et~al\mbox{.}(2016)]%
        {kiene2016surviving}
\bibfield{author}{\bibinfo{person}{Charles Kiene}, \bibinfo{person}{Andr{\'e}s Monroy-Hern{\'a}ndez}, {and} \bibinfo{person}{Benjamin~Mako Hill}.} \bibinfo{year}{2016}\natexlab{}.
\newblock \showarticletitle{Surviving an" eternal september" how an online community managed a surge of newcomers}. In \bibinfo{booktitle}{\emph{Proceedings of the 2016 CHI Conference on Human Factors in Computing Systems}}. \bibinfo{pages}{1152--1156}.
\newblock


\bibitem[Kim et~al\mbox{.}(2016)]%
        {kim2016understanding}
\bibfield{author}{\bibinfo{person}{Suin Kim}, \bibinfo{person}{Sungjoon Park}, \bibinfo{person}{Scott~A Hale}, \bibinfo{person}{Sooyoung Kim}, \bibinfo{person}{Jeongmin Byun}, {and} \bibinfo{person}{Alice~H Oh}.} \bibinfo{year}{2016}\natexlab{}.
\newblock \showarticletitle{Understanding editing behaviors in multilingual Wikipedia}.
\newblock \bibinfo{journal}{\emph{PloS one}} \bibinfo{volume}{11}, \bibinfo{number}{5} (\bibinfo{year}{2016}), \bibinfo{pages}{e0155305}.
\newblock


\bibitem[Kim et~al\mbox{.}(2024)]%
        {kim2024diarymate}
\bibfield{author}{\bibinfo{person}{Taewan Kim}, \bibinfo{person}{Donghoon Shin}, \bibinfo{person}{Young-Ho Kim}, {and} \bibinfo{person}{Hwajung Hong}.} \bibinfo{year}{2024}\natexlab{}.
\newblock \showarticletitle{DiaryMate: Understanding User Perceptions and Experience in Human-AI Collaboration for Personal Journaling}. In \bibinfo{booktitle}{\emph{Proceedings of the 2024 CHI Conference on Human Factors in Computing Systems}}. \bibinfo{pages}{1--15}.
\newblock


\bibitem[Kittur and Kraut(2008)]%
        {kittur2008harnessing}
\bibfield{author}{\bibinfo{person}{Aniket Kittur} {and} \bibinfo{person}{Robert~E Kraut}.} \bibinfo{year}{2008}\natexlab{}.
\newblock \showarticletitle{Harnessing the wisdom of crowds in wikipedia: quality through coordination}. In \bibinfo{booktitle}{\emph{Proceedings of the 2008 ACM conference on Computer supported cooperative work}}. \bibinfo{pages}{37--46}.
\newblock


\bibitem[Kittur and Kraut(2010)]%
        {kittur2010beyond}
\bibfield{author}{\bibinfo{person}{Aniket Kittur} {and} \bibinfo{person}{Robert~E Kraut}.} \bibinfo{year}{2010}\natexlab{}.
\newblock \showarticletitle{Beyond Wikipedia: coordination and conflict in online production groups}. In \bibinfo{booktitle}{\emph{Proceedings of the 2010 ACM conference on Computer supported cooperative work}}. \bibinfo{pages}{215--224}.
\newblock


\bibitem[Kobak et~al\mbox{.}(2024)]%
        {kobak2024delving}
\bibfield{author}{\bibinfo{person}{Dmitry Kobak}, \bibinfo{person}{Rita Gonz{\'a}lez-M{\'a}rquez}, \bibinfo{person}{Em{\H{o}}ke-{\'A}gnes Horv{\'a}t}, {and} \bibinfo{person}{Jan Lause}.} \bibinfo{year}{2024}\natexlab{}.
\newblock \showarticletitle{Delving into ChatGPT usage in academic writing through excess vocabulary}.
\newblock \bibinfo{journal}{\emph{arXiv preprint arXiv:2406.07016}} (\bibinfo{year}{2024}).
\newblock


\bibitem[Kumarana et~al\mbox{.}(2011)]%
        {kumarana2011wikibhasha}
\bibfield{author}{\bibinfo{person}{Narend Kumarana}, \bibinfo{person}{S Ashwani}, {and} \bibinfo{person}{D Vikram}.} \bibinfo{year}{2011}\natexlab{}.
\newblock \showarticletitle{Wikibhasha: Our experiences with multilingual content creation tool for wikipedia}. In \bibinfo{booktitle}{\emph{Proceedings of Wikipedia Conference India, Wikimedia Foundation}}.
\newblock


\bibitem[Kuo et~al\mbox{.}(2024)]%
        {kuo2024wikibench}
\bibfield{author}{\bibinfo{person}{Tzu-Sheng Kuo}, \bibinfo{person}{Aaron~Lee Halfaker}, \bibinfo{person}{Zirui Cheng}, \bibinfo{person}{Jiwoo Kim}, \bibinfo{person}{Meng-Hsin Wu}, \bibinfo{person}{Tongshuang Wu}, \bibinfo{person}{Kenneth Holstein}, {and} \bibinfo{person}{Haiyi Zhu}.} \bibinfo{year}{2024}\natexlab{}.
\newblock \showarticletitle{Wikibench: Community-driven data curation for ai evaluation on wikipedia}. In \bibinfo{booktitle}{\emph{Proceedings of the 2024 CHI Conference on Human Factors in Computing Systems}}. \bibinfo{pages}{1--24}.
\newblock


\bibitem[Lave and Wenger(1991)]%
        {lave1991situated}
\bibfield{author}{\bibinfo{person}{Jean Lave} {and} \bibinfo{person}{Etienne Wenger}.} \bibinfo{year}{1991}\natexlab{}.
\newblock \bibinfo{booktitle}{\emph{Situated learning: Legitimate peripheral participation}}.
\newblock \bibinfo{publisher}{Cambridge university press}.
\newblock


\bibitem[Laxstr{\"o}m et~al\mbox{.}(2015)]%
        {laxstrom2015content}
\bibfield{author}{\bibinfo{person}{Niklas Laxstr{\"o}m}, \bibinfo{person}{Pau Giner}, {and} \bibinfo{person}{Santhosh Thottingal}.} \bibinfo{year}{2015}\natexlab{}.
\newblock \showarticletitle{Content Translation: Computer-assisted translation tool for Wikipedia articles}.
\newblock \bibinfo{journal}{\emph{arXiv preprint arXiv:1506.01914}} (\bibinfo{year}{2015}).
\newblock


\bibitem[Lee et~al\mbox{.}(2024)]%
        {lee2024design}
\bibfield{author}{\bibinfo{person}{Mina Lee}, \bibinfo{person}{Katy~Ilonka Gero}, \bibinfo{person}{John Joon~Young Chung}, \bibinfo{person}{Simon~Buckingham Shum}, \bibinfo{person}{Vipul Raheja}, \bibinfo{person}{Hua Shen}, \bibinfo{person}{Subhashini Venugopalan}, \bibinfo{person}{Thiemo Wambsganss}, \bibinfo{person}{David Zhou}, \bibinfo{person}{Emad~A Alghamdi}, {et~al\mbox{.}}} \bibinfo{year}{2024}\natexlab{}.
\newblock \showarticletitle{A design space for intelligent and interactive writing assistants}. In \bibinfo{booktitle}{\emph{Proceedings of the 2024 CHI Conference on Human Factors in Computing Systems}}. \bibinfo{pages}{1--35}.
\newblock


\bibitem[Lee et~al\mbox{.}(2022)]%
        {lee2022coauthor}
\bibfield{author}{\bibinfo{person}{Mina Lee}, \bibinfo{person}{Percy Liang}, {and} \bibinfo{person}{Qian Yang}.} \bibinfo{year}{2022}\natexlab{}.
\newblock \showarticletitle{Coauthor: Designing a human-ai collaborative writing dataset for exploring language model capabilities}. In \bibinfo{booktitle}{\emph{Proceedings of the 2022 CHI conference on human factors in computing systems}}. \bibinfo{pages}{1--19}.
\newblock


\bibitem[Liu and Ram(2011)]%
        {liu2011does}
\bibfield{author}{\bibinfo{person}{Jun Liu} {and} \bibinfo{person}{Sudha Ram}.} \bibinfo{year}{2011}\natexlab{}.
\newblock \showarticletitle{Who does what: Collaboration patterns in the Wikipedia and their impact on article quality}.
\newblock \bibinfo{journal}{\emph{ACM Transactions on Management Information Systems (TMIS)}} \bibinfo{volume}{2}, \bibinfo{number}{2} (\bibinfo{year}{2011}), \bibinfo{pages}{1--23}.
\newblock


\bibitem[Luther et~al\mbox{.}(2024)]%
        {luther2024teaming}
\bibfield{author}{\bibinfo{person}{Teresa Luther}, \bibinfo{person}{Joachim Kimmerle}, {and} \bibinfo{person}{Ulrike Cress}.} \bibinfo{year}{2024}\natexlab{}.
\newblock \showarticletitle{Teaming up with an AI: Exploring human--AI collaboration in a writing scenario with ChatGPT}.
\newblock \bibinfo{journal}{\emph{AI}} \bibinfo{volume}{5}, \bibinfo{number}{3} (\bibinfo{year}{2024}), \bibinfo{pages}{1357--1376}.
\newblock


\bibitem[MacKenzie(2024)]%
        {mackenzie2024human}
\bibfield{author}{\bibinfo{person}{I~Scott MacKenzie}.} \bibinfo{year}{2024}\natexlab{}.
\newblock \bibinfo{booktitle}{\emph{Human-computer interaction: An empirical research perspective}}.
\newblock \bibinfo{publisher}{Elsevier Science}.
\newblock


\bibitem[Maddock et~al\mbox{.}(2017)]%
        {maddock2017talking}
\bibfield{author}{\bibinfo{person}{Jim Maddock}, \bibinfo{person}{Aaron Shaw}, {and} \bibinfo{person}{Darren Gergle}.} \bibinfo{year}{2017}\natexlab{}.
\newblock \showarticletitle{Talking about talk: coordination in large online communities}. In \bibinfo{booktitle}{\emph{Proceedings of the 2017 CHI Conference Extended Abstracts on Human Factors in Computing Systems}}. \bibinfo{pages}{1869--1876}.
\newblock


\bibitem[Morgan et~al\mbox{.}(2013)]%
        {morgan2013tea}
\bibfield{author}{\bibinfo{person}{Jonathan~T Morgan}, \bibinfo{person}{Siko Bouterse}, \bibinfo{person}{Heather Walls}, {and} \bibinfo{person}{Sarah Stierch}.} \bibinfo{year}{2013}\natexlab{}.
\newblock \showarticletitle{Tea and sympathy: crafting positive new user experiences on wikipedia}. In \bibinfo{booktitle}{\emph{Proceedings of the 2013 conference on Computer supported cooperative work}}. \bibinfo{pages}{839--848}.
\newblock


\bibitem[Morgan and Halfaker(2018)]%
        {morgan2018evaluating}
\bibfield{author}{\bibinfo{person}{Jonathan~T Morgan} {and} \bibinfo{person}{Aaron Halfaker}.} \bibinfo{year}{2018}\natexlab{}.
\newblock \showarticletitle{Evaluating the impact of the Wikipedia Teahouse on newcomer socialization and retention}. In \bibinfo{booktitle}{\emph{Proceedings of the 14th international symposium on open collaboration}}. \bibinfo{pages}{1--7}.
\newblock


\bibitem[Nguyen et~al\mbox{.}(2024)]%
        {nguyen2024human}
\bibfield{author}{\bibinfo{person}{Andy Nguyen}, \bibinfo{person}{Yvonne Hong}, \bibinfo{person}{Belle Dang}, {and} \bibinfo{person}{Xiaoshan Huang}.} \bibinfo{year}{2024}\natexlab{}.
\newblock \showarticletitle{Human-AI collaboration patterns in AI-assisted academic writing}.
\newblock \bibinfo{journal}{\emph{Studies in Higher Education}} \bibinfo{volume}{49}, \bibinfo{number}{5} (\bibinfo{year}{2024}), \bibinfo{pages}{847--864}.
\newblock


\bibitem[Niederer and Van~Dijck(2010)]%
        {niederer2010wisdom}
\bibfield{author}{\bibinfo{person}{Sabine Niederer} {and} \bibinfo{person}{Jos{\'e} Van~Dijck}.} \bibinfo{year}{2010}\natexlab{}.
\newblock \showarticletitle{Wisdom of the crowd or technicity of content? Wikipedia as a sociotechnical system}.
\newblock \bibinfo{journal}{\emph{New media \& society}} \bibinfo{volume}{12}, \bibinfo{number}{8} (\bibinfo{year}{2010}), \bibinfo{pages}{1368--1387}.
\newblock


\bibitem[Norman(1991)]%
        {norman1991cognitive}
\bibfield{author}{\bibinfo{person}{Donald~A Norman}.} \bibinfo{year}{1991}\natexlab{}.
\newblock \showarticletitle{Cognitive artifacts}.
\newblock \bibinfo{journal}{\emph{Designing interaction: Psychology at the human-computer interface}} \bibinfo{volume}{1}, \bibinfo{number}{1} (\bibinfo{year}{1991}), \bibinfo{pages}{17--38}.
\newblock


\bibitem[Oulasvirta and Hornb{\ae}k(2016)]%
        {oulasvirta2016hci}
\bibfield{author}{\bibinfo{person}{Antti Oulasvirta} {and} \bibinfo{person}{Kasper Hornb{\ae}k}.} \bibinfo{year}{2016}\natexlab{}.
\newblock \showarticletitle{HCI research as problem-solving}. In \bibinfo{booktitle}{\emph{Proceedings of the 2016 CHI Conference on Human Factors in Computing Systems}}. \bibinfo{pages}{4956--4967}.
\newblock


\bibitem[Preece and Shneiderman(2009)]%
        {preece2009reader}
\bibfield{author}{\bibinfo{person}{Jennifer Preece} {and} \bibinfo{person}{Ben Shneiderman}.} \bibinfo{year}{2009}\natexlab{}.
\newblock \showarticletitle{The reader-to-leader framework: Motivating technology-mediated social participation}.
\newblock \bibinfo{journal}{\emph{AIS transactions on human-computer interaction}} \bibinfo{volume}{1}, \bibinfo{number}{1} (\bibinfo{year}{2009}), \bibinfo{pages}{13--32}.
\newblock


\bibitem[Priedhorsky et~al\mbox{.}(2007)]%
        {priedhorsky2007creating}
\bibfield{author}{\bibinfo{person}{Reid Priedhorsky}, \bibinfo{person}{Jilin Chen}, \bibinfo{person}{Shyong (Tony)~K Lam}, \bibinfo{person}{Katherine Panciera}, \bibinfo{person}{Loren Terveen}, {and} \bibinfo{person}{John Riedl}.} \bibinfo{year}{2007}\natexlab{}.
\newblock \showarticletitle{Creating, destroying, and restoring value in Wikipedia}. In \bibinfo{booktitle}{\emph{Proceedings of the 2007 ACM international conference on supporting group work}}. \bibinfo{pages}{259--268}.
\newblock


\bibitem[Radensky et~al\mbox{.}(2024)]%
        {radensky2024let}
\bibfield{author}{\bibinfo{person}{Marissa Radensky}, \bibinfo{person}{Daniel~S Weld}, \bibinfo{person}{Joseph~Chee Chang}, \bibinfo{person}{Pao Siangliulue}, {and} \bibinfo{person}{Jonathan Bragg}.} \bibinfo{year}{2024}\natexlab{}.
\newblock \showarticletitle{Let's Get to the Point: LLM-Supported Planning, Drafting, and Revising of Research-Paper Blog Posts}.
\newblock \bibinfo{journal}{\emph{arXiv preprint arXiv:2406.10370}} (\bibinfo{year}{2024}).
\newblock


\bibitem[Rajagopal et~al\mbox{.}(2022)]%
        {rajagopal2022one}
\bibfield{author}{\bibinfo{person}{Dheeraj Rajagopal}, \bibinfo{person}{Xuchao Zhang}, \bibinfo{person}{Michael Gamon}, \bibinfo{person}{Sujay~Kumar Jauhar}, \bibinfo{person}{Diyi Yang}, {and} \bibinfo{person}{Eduard Hovy}.} \bibinfo{year}{2022}\natexlab{}.
\newblock \showarticletitle{One document, many revisions: A dataset for classification and description of edit intents}. In \bibinfo{booktitle}{\emph{Proceedings of the Thirteenth Language Resources and Evaluation Conference}}. \bibinfo{pages}{5517--5524}.
\newblock


\bibitem[Redi et~al\mbox{.}(2020)]%
        {redi2020taxonomy}
\bibfield{author}{\bibinfo{person}{Miriam Redi}, \bibinfo{person}{Martin Gerlach}, \bibinfo{person}{Isaac Johnson}, \bibinfo{person}{Jonathan Morgan}, {and} \bibinfo{person}{Leila Zia}.} \bibinfo{year}{2020}\natexlab{}.
\newblock \showarticletitle{A taxonomy of knowledge gaps for wikimedia projects (second draft)}.
\newblock \bibinfo{journal}{\emph{arXiv preprint arXiv:2008.12314}} (\bibinfo{year}{2020}).
\newblock


\bibitem[Ren et~al\mbox{.}(2023)]%
        {ren2023did}
\bibfield{author}{\bibinfo{person}{Yuqing Ren}, \bibinfo{person}{Haifeng Zhang}, {and} \bibinfo{person}{Robert~E Kraut}.} \bibinfo{year}{2023}\natexlab{}.
\newblock \showarticletitle{How did they build the free encyclopedia? a literature review of collaboration and coordination among Wikipedia editors}.
\newblock \bibinfo{journal}{\emph{ACM Transactions on Computer-Human Interaction}} \bibinfo{volume}{31}, \bibinfo{number}{1} (\bibinfo{year}{2023}), \bibinfo{pages}{1--48}.
\newblock


\bibitem[Roberts(2014)]%
        {roberts2014community}
\bibfield{author}{\bibinfo{person}{Joanne Roberts}.} \bibinfo{year}{2014}\natexlab{}.
\newblock \showarticletitle{Community and the dynamics of spatially distributed knowledge production: the case of Wikipedia}.
\newblock In \bibinfo{booktitle}{\emph{The social dynamics of innovation networks}}. \bibinfo{publisher}{Routledge}, \bibinfo{pages}{179--200}.
\newblock


\bibitem[Roy et~al\mbox{.}(2022)]%
        {roy2022information}
\bibfield{author}{\bibinfo{person}{Dwaipayan Roy}, \bibinfo{person}{Sumit Bhatia}, {and} \bibinfo{person}{Prateek Jain}.} \bibinfo{year}{2022}\natexlab{}.
\newblock \showarticletitle{Information asymmetry in Wikipedia across different languages: A statistical analysis}.
\newblock \bibinfo{journal}{\emph{Journal of the Association for Information Science and Technology}} \bibinfo{volume}{73}, \bibinfo{number}{3} (\bibinfo{year}{2022}), \bibinfo{pages}{347--361}.
\newblock


\bibitem[Ruprechter et~al\mbox{.}(2020a)]%
        {ruprechter2020relation}
\bibfield{author}{\bibinfo{person}{Thorsten Ruprechter}, \bibinfo{person}{Tiago Santos}, {and} \bibinfo{person}{Denis Helic}.} \bibinfo{year}{2020}\natexlab{a}.
\newblock \showarticletitle{On the relation of edit behavior, link structure, and article quality on wikipedia}. In \bibinfo{booktitle}{\emph{Complex Networks and Their Applications VIII: Volume 2 Proceedings of the Eighth International Conference on Complex Networks and Their Applications COMPLEX NETWORKS 2019 8}}. Springer, \bibinfo{pages}{242--254}.
\newblock


\bibitem[Ruprechter et~al\mbox{.}(2020b)]%
        {ruprechter2020relating}
\bibfield{author}{\bibinfo{person}{Thorsten Ruprechter}, \bibinfo{person}{Tiago Santos}, {and} \bibinfo{person}{Denis Helic}.} \bibinfo{year}{2020}\natexlab{b}.
\newblock \showarticletitle{Relating Wikipedia article quality to edit behavior and link structure}.
\newblock \bibinfo{journal}{\emph{Applied Network Science}}  \bibinfo{volume}{5} (\bibinfo{year}{2020}), \bibinfo{pages}{1--20}.
\newblock


\bibitem[{\v{S}}akota et~al\mbox{.}(2024)]%
        {vsakota2024edisum}
\bibfield{author}{\bibinfo{person}{Marija {\v{S}}akota}, \bibinfo{person}{Isaac Johnson}, \bibinfo{person}{Guosheng Feng}, {and} \bibinfo{person}{Robert West}.} \bibinfo{year}{2024}\natexlab{}.
\newblock \showarticletitle{Edisum: Summarizing and explaining wikipedia edits at scale}.
\newblock \bibinfo{journal}{\emph{arXiv e-prints}} (\bibinfo{year}{2024}), \bibinfo{pages}{arXiv--2404}.
\newblock


\bibitem[Shao et~al\mbox{.}(2024)]%
        {shao2024assisting}
\bibfield{author}{\bibinfo{person}{Yijia Shao}, \bibinfo{person}{Yucheng Jiang}, \bibinfo{person}{Theodore~A Kanell}, \bibinfo{person}{Peter Xu}, \bibinfo{person}{Omar Khattab}, {and} \bibinfo{person}{Monica~S Lam}.} \bibinfo{year}{2024}\natexlab{}.
\newblock \showarticletitle{Assisting in writing wikipedia-like articles from scratch with large language models}.
\newblock \bibinfo{journal}{\emph{arXiv preprint arXiv:2402.14207}} (\bibinfo{year}{2024}).
\newblock


\bibitem[Smith et~al\mbox{.}(2020)]%
        {smith2020keeping}
\bibfield{author}{\bibinfo{person}{C~Estelle Smith}, \bibinfo{person}{Bowen Yu}, \bibinfo{person}{Anjali Srivastava}, \bibinfo{person}{Aaron Halfaker}, \bibinfo{person}{Loren Terveen}, {and} \bibinfo{person}{Haiyi Zhu}.} \bibinfo{year}{2020}\natexlab{}.
\newblock \showarticletitle{Keeping community in the loop: Understanding wikipedia stakeholder values for machine learning-based systems}. In \bibinfo{booktitle}{\emph{Proceedings of the 2020 CHI Conference on Human Factors in Computing Systems}}. \bibinfo{pages}{1--14}.
\newblock


\bibitem[Stefnisson and Thue(2018)]%
        {stefnisson2018mimisbrunnur}
\bibfield{author}{\bibinfo{person}{Ingibergur Stefnisson} {and} \bibinfo{person}{David Thue}.} \bibinfo{year}{2018}\natexlab{}.
\newblock \showarticletitle{Mimisbrunnur: AI-assisted authoring for interactive storytelling}. In \bibinfo{booktitle}{\emph{Proceedings of the AAAI Conference on artificial Intelligence and Interactive Digital entertainment}}, Vol.~\bibinfo{volume}{14}. \bibinfo{pages}{236--242}.
\newblock


\bibitem[Steinmacher et~al\mbox{.}(2015)]%
        {steinmacher2015systematic}
\bibfield{author}{\bibinfo{person}{Igor Steinmacher}, \bibinfo{person}{Marco Aurelio~Graciotto Silva}, \bibinfo{person}{Marco~Aurelio Gerosa}, {and} \bibinfo{person}{David~F Redmiles}.} \bibinfo{year}{2015}\natexlab{}.
\newblock \showarticletitle{A systematic literature review on the barriers faced by newcomers to open source software projects}.
\newblock \bibinfo{journal}{\emph{Information and Software Technology}}  \bibinfo{volume}{59} (\bibinfo{year}{2015}), \bibinfo{pages}{67--85}.
\newblock


\bibitem[Suh et~al\mbox{.}(2009)]%
        {suh2009singularity}
\bibfield{author}{\bibinfo{person}{Bongwon Suh}, \bibinfo{person}{Gregorio Convertino}, \bibinfo{person}{Ed~H Chi}, {and} \bibinfo{person}{Peter Pirolli}.} \bibinfo{year}{2009}\natexlab{}.
\newblock \showarticletitle{The singularity is not near: slowing growth of Wikipedia}. In \bibinfo{booktitle}{\emph{Proceedings of the 5th international symposium on wikis and open collaboration}}. \bibinfo{pages}{1--10}.
\newblock


\bibitem[Tang et~al\mbox{.}(2024)]%
        {tang2024llms}
\bibfield{author}{\bibinfo{person}{Xuemei Tang}, \bibinfo{person}{Xufeng Duan}, {and} \bibinfo{person}{Zhenguang~G Cai}.} \bibinfo{year}{2024}\natexlab{}.
\newblock \showarticletitle{Are LLMs Good Literature Review Writers? Evaluating the Literature Review Writing Ability of Large Language Models}.
\newblock \bibinfo{journal}{\emph{arXiv preprint arXiv:2412.13612}} (\bibinfo{year}{2024}).
\newblock


\bibitem[Tang et~al\mbox{.}(2025)]%
        {tang2025understanding}
\bibfield{author}{\bibinfo{person}{Yuying Tang}, \bibinfo{person}{Haotian Li}, \bibinfo{person}{Minghe Lan}, \bibinfo{person}{Xiaojuan Ma}, {and} \bibinfo{person}{Huamin Qu}.} \bibinfo{year}{2025}\natexlab{}.
\newblock \showarticletitle{Understanding Screenwriters' Practices, Attitudes, and Future Expectations in Human-AI Co-Creation}.
\newblock \bibinfo{journal}{\emph{arXiv preprint arXiv:2502.16153}} (\bibinfo{year}{2025}).
\newblock


\bibitem[Tsvetkova et~al\mbox{.}(2017)]%
        {tsvetkova2017even}
\bibfield{author}{\bibinfo{person}{Milena Tsvetkova}, \bibinfo{person}{Ruth Garc{\'\i}a-Gavilanes}, \bibinfo{person}{Luciano Floridi}, {and} \bibinfo{person}{Taha Yasseri}.} \bibinfo{year}{2017}\natexlab{}.
\newblock \showarticletitle{Even good bots fight: The case of Wikipedia}.
\newblock \bibinfo{journal}{\emph{PloS one}} \bibinfo{volume}{12}, \bibinfo{number}{2} (\bibinfo{year}{2017}), \bibinfo{pages}{e0171774}.
\newblock


\bibitem[Velt et~al\mbox{.}(2020)]%
        {velt2020translations}
\bibfield{author}{\bibinfo{person}{Raphael Velt}, \bibinfo{person}{Steve Benford}, {and} \bibinfo{person}{Stuart Reeves}.} \bibinfo{year}{2020}\natexlab{}.
\newblock \showarticletitle{Translations and boundaries in the gap between HCI theory and design practice}.
\newblock \bibinfo{journal}{\emph{ACM Transactions on Computer-Human Interaction (TOCHI)}} \bibinfo{volume}{27}, \bibinfo{number}{4} (\bibinfo{year}{2020}), \bibinfo{pages}{1--28}.
\newblock


\bibitem[Wasi et~al\mbox{.}(2024)]%
        {wasi2024llms}
\bibfield{author}{\bibinfo{person}{Azmine~Toushik Wasi}, \bibinfo{person}{Mst~Rafia Islam}, {and} \bibinfo{person}{Raima Islam}.} \bibinfo{year}{2024}\natexlab{}.
\newblock \showarticletitle{Llms as writing assistants: Exploring perspectives on sense of ownership and reasoning}. In \bibinfo{booktitle}{\emph{Proceedings of the Third Workshop on Intelligent and Interactive Writing Assistants}}. \bibinfo{pages}{38--42}.
\newblock


\bibitem[{Wikimedia}(2025)]%
        {wikimedia}
\bibfield{author}{\bibinfo{person}{{Wikimedia}}.} \bibinfo{year}{2025}\natexlab{}.
\newblock \bibinfo{title}{Strategy/Multigenerational/Artificial intelligence for editors}.
\newblock
\urldef\tempurl%
\url{https://meta.m.wikimedia.org/wiki/Strategy/Multigenerational/Artificial_intelligence_for_editors}
\showURL{%
\tempurl}
\newblock
\shownote{Accessed: 2025-05-06}.


\bibitem[{Wikipedia}(2024a)]%
        {wikipedia_npoV}
\bibfield{author}{\bibinfo{person}{{Wikipedia}}.} \bibinfo{year}{2024}\natexlab{a}.
\newblock \bibinfo{title}{Wikipedia:Neutral point of view}.
\newblock
\urldef\tempurl%
\url{https://en.wikipedia.org/wiki/Wikipedia:Neutral_point_of_view}
\showURL{%
\tempurl}
\newblock
\shownote{Accessed: 2025-04-24}.


\bibitem[{Wikipedia}(2024b)]%
        {npov}
\bibfield{author}{\bibinfo{person}{{Wikipedia}}.} \bibinfo{year}{2024}\natexlab{b}.
\newblock \bibinfo{title}{Wikipedia:Neutral point of view}.
\newblock
\urldef\tempurl%
\url{https://en.wikipedia.org/wiki/Wikipedia:Neutral_point_of_view}
\showURL{%
\tempurl}
\newblock
\shownote{Accessed: 2025-04-30}.


\bibitem[{Wikipedia}(2024c)]%
        {nor}
\bibfield{author}{\bibinfo{person}{{Wikipedia}}.} \bibinfo{year}{2024}\natexlab{c}.
\newblock \bibinfo{title}{Wikipedia:No original research}.
\newblock
\urldef\tempurl%
\url{https://en.wikipedia.org/wiki/Wikipedia:No_original_research}
\showURL{%
\tempurl}
\newblock
\shownote{Accessed: 2025-04-30}.


\bibitem[{Wikipedia}(2024d)]%
        {wikipedia_notability}
\bibfield{author}{\bibinfo{person}{{Wikipedia}}.} \bibinfo{year}{2024}\natexlab{d}.
\newblock \bibinfo{title}{Wikipedia:Notability}.
\newblock
\urldef\tempurl%
\url{https://en.wikipedia.org/wiki/Wikipedia:Notability}
\showURL{%
\tempurl}
\newblock
\shownote{Accessed: 2025-04-24}.


\bibitem[{Wikipedia}(2024e)]%
        {wikipedia_verifiability}
\bibfield{author}{\bibinfo{person}{{Wikipedia}}.} \bibinfo{year}{2024}\natexlab{e}.
\newblock \bibinfo{title}{Wikipedia:Verifiability}.
\newblock
\urldef\tempurl%
\url{https://en.wikipedia.org/wiki/Wikipedia:Verifiability}
\showURL{%
\tempurl}
\newblock
\shownote{Accessed: 2025-04-24}.


\bibitem[{Wikipedia}(2024f)]%
        {verifiability}
\bibfield{author}{\bibinfo{person}{{Wikipedia}}.} \bibinfo{year}{2024}\natexlab{f}.
\newblock \bibinfo{title}{Wikipedia:Verifiability}.
\newblock
\urldef\tempurl%
\url{https://en.wikipedia.org/wiki/Wikipedia:Verifiability}
\showURL{%
\tempurl}
\newblock
\shownote{Accessed: 2025-04-30}.


\bibitem[Wulczyn et~al\mbox{.}(2016)]%
        {wulczyn2016growing}
\bibfield{author}{\bibinfo{person}{Ellery Wulczyn}, \bibinfo{person}{Robert West}, \bibinfo{person}{Leila Zia}, {and} \bibinfo{person}{Jure Leskovec}.} \bibinfo{year}{2016}\natexlab{}.
\newblock \showarticletitle{Growing wikipedia across languages via recommendation}. In \bibinfo{booktitle}{\emph{Proceedings of the 25th International Conference on World Wide Web}}. \bibinfo{pages}{975--985}.
\newblock


\bibitem[Yang et~al\mbox{.}(2016a)]%
        {yang2016edit}
\bibfield{author}{\bibinfo{person}{Diyi Yang}, \bibinfo{person}{Aaron Halfaker}, \bibinfo{person}{Robert Kraut}, {and} \bibinfo{person}{Eduard Hovy}.} \bibinfo{year}{2016}\natexlab{a}.
\newblock \showarticletitle{Edit categories and editor role identification in Wikipedia}. In \bibinfo{booktitle}{\emph{Proceedings of the Tenth International Conference on Language Resources and Evaluation (LREC'16)}}. \bibinfo{pages}{1295--1299}.
\newblock


\bibitem[Yang et~al\mbox{.}(2016b)]%
        {yang2016did}
\bibfield{author}{\bibinfo{person}{Diyi Yang}, \bibinfo{person}{Aaron Halfaker}, \bibinfo{person}{Robert Kraut}, {and} \bibinfo{person}{Eduard Hovy}.} \bibinfo{year}{2016}\natexlab{b}.
\newblock \showarticletitle{Who did what: Editor role identification in Wikipedia}. In \bibinfo{booktitle}{\emph{Proceedings of the international AAAI conference on web and social media}}, Vol.~\bibinfo{volume}{10}. \bibinfo{pages}{446--455}.
\newblock


\bibitem[Yang et~al\mbox{.}(2022)]%
        {yang2022ai}
\bibfield{author}{\bibinfo{person}{Daijin Yang}, \bibinfo{person}{Yanpeng Zhou}, \bibinfo{person}{Zhiyuan Zhang}, \bibinfo{person}{Toby Jia-Jun Li}, {and} \bibinfo{person}{Ray Lc}.} \bibinfo{year}{2022}\natexlab{}.
\newblock \showarticletitle{AI as an Active Writer: Interaction strategies with generated text in human-AI collaborative fiction writing}. In \bibinfo{booktitle}{\emph{Joint Proceedings of the ACM IUI Workshops}}, Vol.~\bibinfo{volume}{10}. CEUR-WS Team, \bibinfo{pages}{1--11}.
\newblock


\bibitem[Yang et~al\mbox{.}(2025)]%
        {yang2025modifying}
\bibfield{author}{\bibinfo{person}{Kaixun Yang}, \bibinfo{person}{Mladen Rakovi{\'c}}, \bibinfo{person}{Zhiping Liang}, \bibinfo{person}{Lixiang Yan}, \bibinfo{person}{Zijie Zeng}, \bibinfo{person}{Yizhou Fan}, \bibinfo{person}{Dragan Ga{\v{s}}evi{\'c}}, {and} \bibinfo{person}{Guanliang Chen}.} \bibinfo{year}{2025}\natexlab{}.
\newblock \showarticletitle{Modifying AI, enhancing essays: How active engagement with generative AI boosts writing quality}. In \bibinfo{booktitle}{\emph{Proceedings of the 15th International Learning Analytics and Knowledge Conference}}. \bibinfo{pages}{568--578}.
\newblock


\bibitem[Yasseri et~al\mbox{.}(2012)]%
        {yasseri2012dynamics}
\bibfield{author}{\bibinfo{person}{Taha Yasseri}, \bibinfo{person}{Robert Sumi}, \bibinfo{person}{Andr{\'a}s Rung}, \bibinfo{person}{Andr{\'a}s Kornai}, {and} \bibinfo{person}{J{\'a}nos Kert{\'e}sz}.} \bibinfo{year}{2012}\natexlab{}.
\newblock \showarticletitle{Dynamics of conflicts in Wikipedia}.
\newblock \bibinfo{journal}{\emph{PloS one}} \bibinfo{volume}{7}, \bibinfo{number}{6} (\bibinfo{year}{2012}), \bibinfo{pages}{e38869}.
\newblock


\bibitem[Yuan et~al\mbox{.}(2022)]%
        {yuan2022wordcraft}
\bibfield{author}{\bibinfo{person}{Ann Yuan}, \bibinfo{person}{Andy Coenen}, \bibinfo{person}{Emily Reif}, {and} \bibinfo{person}{Daphne Ippolito}.} \bibinfo{year}{2022}\natexlab{}.
\newblock \showarticletitle{Wordcraft: story writing with large language models}. In \bibinfo{booktitle}{\emph{Proceedings of the 27th International Conference on Intelligent User Interfaces}}. \bibinfo{pages}{841--852}.
\newblock


\bibitem[Zheng et~al\mbox{.}(2019)]%
        {zheng2019roles}
\bibfield{author}{\bibinfo{person}{Lei Zheng}, \bibinfo{person}{Christopher~M Albano}, \bibinfo{person}{Neev~M Vora}, \bibinfo{person}{Feng Mai}, {and} \bibinfo{person}{Jeffrey~V Nickerson}.} \bibinfo{year}{2019}\natexlab{}.
\newblock \showarticletitle{The roles bots play in Wikipedia}.
\newblock \bibinfo{journal}{\emph{Proceedings of the ACM on Human-Computer Interaction}} \bibinfo{volume}{3}, \bibinfo{number}{CSCW} (\bibinfo{year}{2019}), \bibinfo{pages}{1--20}.
\newblock


\end{thebibliography}

\end{document}